\documentclass[preprint,showpacs,preprintnumbers,amsmath,amssymb]{revtex4-1}
\usepackage{graphics,epsfig,subfigure}
\usepackage{diagbox}
\usepackage[usenames]{color}
\usepackage[colorlinks,
linkcolor=blue,
anchorcolor=red,
citecolor=red
]{hyperref}
\usepackage{color}
\usepackage{graphicx}
\usepackage{amsmath}
\usepackage{amsfonts}
\usepackage{amssymb}
\usepackage{indentfirst}
\usepackage{booktabs}

\begin{document}
	\renewcommand{\baselinestretch}{1.3}
	\newcommand\beq{\begin{equation}}
		\newcommand\eeq{\end{equation}}
	\newcommand\beqn{\begin{eqnarray}}
		\newcommand\eeqn{\end{eqnarray}}
	\newcommand\nn{\nonumber}
	\newcommand\fc{\frac}
	\newcommand\lt{\left}
	\newcommand\rt{\right}
	\newcommand\pt{\partial}

\title{\Large{\bf Proca stars and their frozen states  in an infinite tower of higher-derivative gravity}}

\author{ Jun-Ru Chen and  Yong-Qiang Wang\footnote{yqwang@lzu.edu.cn, corresponding author}
	}
	
\affiliation{$^{1}$ Lanzhou Center for Theoretical Physics, Key Laboratory of Theoretical Physics of Gansu Province, School of Physical Science and Technology, Lanzhou University, Lanzhou 730000, China\\
 $^{2}$ Institute of Theoretical Physics $\&$ Research Center of Gravitation, Lanzhou University, Lanzhou 730000, China}
	\date{\today}

    \begin{abstract}
 In this work, we investigate the five-dimensional Proca star under gravity with the infinite tower of higher curvature corrections. We find that when the coupling constant exceeds a critical value, solutions with a frequency approaching zero appear. In the finite-order corrections case $n=2$ (Gauss–Bonnet gravity), the matter field and energy density diverge near the origin as $\omega\to 0$.
  In contrast, for $n\geq3$, the divergence is efficiently suppressed, both the field and the energy density remain finite everywhere, and both the matter field and energy density remain finite everywhere. In the limit $\omega \to 0$, a class of horizonless frozen star solutions emerges, which are referred to ``frozen stars". Importantly, frozen stars contain neither curvature singularities nor event horizons. These frozen stars develop a critical horizon at a finite radius $r_c$, where $-g_{tt}$ and $1/g_{rr}$ approach zero. The frozen star is indistinguishable from that of an extremal black hole outside $r_c$, and its compactness can reach the extremal black hole value.

\end{abstract}	

\maketitle
	
     \newpage
	\section{Introduction} 
Black holes, among the most important compact objects predicted by general relativity (GR), have long been a central focus of theoretical physics and astrophysics. 
Since Schwarzschild derived the first static and spherically symmetric black hole solution in 1916~\cite{Schwarzschild:1916uq}, the exterior geometry of black holes predicted by GR has been subjected to extensive observational tests~\cite{Will:2014kxa}.
These tests include the dynamical measurements of the black hole candidate Cygnus X-1~\cite{Orosz:2011np} and the Event Horizon Telescope (EHT) images of the supermassive black holes in M87~\cite{EventHorizonTelescope:2019dse, EventHorizonTelescope:2019uob,
EventHorizonTelescope:2019jan,
EventHorizonTelescope:2019ths,
EventHorizonTelescope:2019pgp,
EventHorizonTelescope:2019ggy} and Sagittarius  A*~\cite{EventHorizonTelescope:2022wkp,
EventHorizonTelescope:2022apq,
EventHorizonTelescope:2022wok,
EventHorizonTelescope:2022exc,
EventHorizonTelescope:2022urf,
EventHorizonTelescope:2022xqj}.
Despite these successes, classical GR predicts that black holes inevitably contain curvature singularities, where curvature diverges and classical physical laws lose their validity.
It is widely expected that quantum gravitational effects may solve the spacetime singularities. 
However, a complete and self-consistent theory of quantum gravity has yet to be established.
The Penrose–Hawking singularity theorems demonstrate that, under certain energy conditions, singularity formation is unavoidable in classical gravity~\cite{Penrose:1964wq,
Hawking:1966sx,
Hawking:1966jv,
Hawking:1967ju}. 

A widely studied approach to the singularity problem is the construction of spacetimes with a regular center, leading to regular black hole (RBH) solutions~\cite{Ansoldi:2008jw}. 
These models retain the Schwarzschild exterior geometry, while the central region is described by a regular core where curvature is finite, thereby removing the classical singularity.
 Early attempts include the study of Schwarzschild–Nordström type solutions without singularities by Shirokov in 1948~\cite{Shirokov1948}.  Subsequently, Duan studied regular solutions of the Einstein–Maxwell equations for point-like charges~\cite{Duan:1954bms}, establishing the foundations for the construction of RBHs. 
 In the 1960s, Sakharov and Gliner further proposed that replacing the spacetime near the center with a de Sitter–like vacuum can eliminate the classical singularity~\cite{Sakharov:1966aja, Gliner:1966cgu}.
 The first well-known RBH model was proposed by Bardeen in 1968~\cite{bardeen1968non}. 
 It features a de Sitter–like core and asymptotically approaches the Schwarzschild solution at large distances. Later, Hayward constructed a more general RBH model in 2006~\cite{Hayward:2005gi}, introducing a length scale that controls the size of its de Sitter core.
 However, these RBH solutions typically rely on exotic matter sources in the central region, such as nonlinear electromagnetic fields, to support the regular core~\cite{Carballo-Rubio:2023mvr}. The absence of a clear microscopic origin for these matter components continues to motivate alternative approaches that resolve the singularity problem through purely geometric modifications of the gravitational sector.

    In $D\geq 5$ dimensions, pure gravity theories with the infinite tower of higher-curvature terms admit vacuum black hole metrics with a regular core~\cite{Bueno:2024dgm}. In these solutions, curvature remains finite at the center, while the large radius spacetime geometry is asymptotically flat and Schwarzschild-like.
    This provides a concrete geometric avenue to address the singularity problem without introducing matter fields.
   Subsequent work has investigated whether a purely gravitational mechanism analogous to this can be realized in four-dimensional spacetime~\cite{Bueno:2025zaj}. These theories are part of the quasi-topological gravity class~\cite{Oliva:2010eb,
Myers:2010ru,
Dehghani:2011vu,
Cisterna:2017umf,
Ahmed:2017jod}. Recent studies have further investigated RBHs in pure gravity, including  quasinormal mode, dynamical formation, holographic properties, and thermodynamics, etc~\cite{Stashko:2024wuq,
Konoplya:2024hfg,
Konoplya:2024kih, Arbelaez:2025gwj, Bueno:2024eig,
Bueno:2024zsx,
Bueno:2025gjg, Caceres:2024edr, Aguayo:2025xfi, Ditta:2024iky,
Wang:2024zlq,
Cisterna:2025vxk,
Hao:2025utc}.

While these RBHs resolve the central singularity, they still possess an event horizon. This motivates the study of nonsingular and horizonless compact objects, with boson stars serving as a primary example. Boson stars are supported by bosonic fields in gravitational equilibrium, forming self-gravitating solitons whose spacetime remains regular throughout.
 This idea can be traced back to Wheeler’s geons, which are self-gravitating objects arising from the nonlinear interaction between electromagnetic fields and gravity~\cite{Wheeler:1955zz,
Power:1957zz}. However, geons are unstable and will collapse under perturbations. Kaup, Ruffini and Bonazzola replaced the electromagnetic fields with matter fields and obtained the first stable, static solutions, referred to as boson stars~\cite{Kaup:1968zz,
Ruffini:1969qy}. Later studies extended boson star models from scalar fields to spin-1 (Proca) and spin-1/2 (Dirac) fields, giving rise to Proca stars~\cite{Brito:2015pxa,
SalazarLandea:2016bys} and Dirac stars~\cite{Finster:1998ws,
Finster:1998ux,
Dzhunushaliev:2018jhj}, respectively. Such compact objects have become increasingly relevant in strong-gravity and gravitational-wave astrophysics, owing to their potential observational signatures~\cite{Liebling:2012fv,
Cardoso:2019rvt}.

Recent work has explored gravity theories with an infinite tower of curvature corrections, minimally coupled to scalar fields, and constructed regular solutions~\cite{Ma:2024olw}.
In this model, when the coupling constant exceeds a certain critical threshold, solutions with frequency approaching zero emerge. These solutions do not possess an event horizon but instead have a critical horizon. Inside the critical horizon, the metric component $-g_{tt}$ tends to zero, indicating the presence of an extremely large gravitational redshift. As a result, to a distant observer, dynamical processes near the critical horizon appear to slow down, and the star appears to be frozen. In this work, we investigate spherically symmetric and static Proca star solutions within higher-derivative gravity theories, analyzing how curvature corrections alter their fundamental properties. Through numerical methods, we obtain a class of spherically symmetric static solutions without the event horizon.
And within a specific parameter region, we find frozen star solutions. 
 From the perspective of observers at infinity, the external geometry of these frozen star solutions is difficult to distinguish from that of extremal black holes.

    The paper is organized as follows. In Section~\ref{sec2}, we introduce the Proca star model in gravity theories with higher-curvature corrections of order $n$. In Section~\ref{sec3}, we provide the numerical methods and discuss the associated boundary conditions. In Section~\ref{sec4}, we present the numerical solutions for correction orders
     $n = 1, 2, 3, 4$ and $\infty$, and analyze their physical properties. Finally, Section~\ref{sec5} is the conclusions and discussion.

       \section{THE MODEL SETUP} \label{sec2}
       We consider a Proca field minimally coupled to a gravity theory with an infinite tower of higher-curvature terms. The high-curvature gravitational theory is constructed from arbitrary contractions of the Riemann tensor and the metric, with the Lagrangian density $\mathcal{L}(g^{ab}, R_{cdef})$. The action of the system is
       \begin{equation}\label{equ1}
	S=\int\mathrm d^Dx \sqrt{|g|} \left[\frac{R}{16\pi G}+\sum_{n=2}^{n_{\text{max}}}\frac{\alpha_n\mathcal{Z}_n}{16\pi G}+\mathcal{L}_P\right] \ , 
\end{equation}
    where $|g|$ is the determinant of the metric tensor $g_{ab}$, $G$ is a gravitational constant, $R$ represents a Ricci scalar, and $\alpha_n$ are arbitrary the coupling parameters. $\mathcal{Z}_n$ is the density of order $n$ in the Riemann tensor, and its specific expression can be found in~\cite{Bueno:2024dgm, Konoplya:2024kih}. The Lagrangian density for the Proca field is 
    \begin{equation}\label{equ2}
	\mathcal{L}_P=- \frac{1}{4} \: \mathcal{F}_{\alpha\beta} \bar{\mathcal{F}}^{\alpha\beta} - \frac{\mu^2}{2} \mathcal{A}_\alpha \bar{\mathcal{A}}^\alpha \ , 
\end{equation}
   where $\mathcal{A}$ represents the Proca field and the field strength $\mathcal{F}=d\mathcal{A}$, $\bar{\mathcal{A}}$ and $\bar{\mathcal{F}}$ represents the corresponding complex conjugate.
For static, spherically symmetric spacetimes, we adopt the metric ansatz
\begin{equation}\label{equ6}
	\mathrm ds^2=-\sigma(r)^2N(r)\mathrm dt^2+\frac{\mathrm dr^2}{N(r)}+r^2\mathrm d\Omega_{D-2}^2 \ ,
\end{equation}
and the Proca field ansatz
\begin{equation}\label{equ7}
	\mathcal{A}=[F(r) d t+i H(r) d r] e^{-i \omega t}
\end{equation}
where $F(r)$ and $H(r)$ are two real potentials, and $\omega$ is the field oscillation frequency.

In four-dimensional spacetime, the Gauss–Bonnet term is dynamically trivial for spherically symmetric configurations. To investigate the impact of higher-curvature corrections on the structure of Proca stars, we consider five-dimensional ($D=5$) spacetime as the minimal nontrivial setting. Substituting the metric ansatz~(\ref{equ6}) and the Proca field ansatz~(\ref{equ7}) into the action~(\ref{equ1}), and varying the action, yields the equations 
\begin{equation}\label{equ8}
 \frac{ H\left(\omega^2-\mu^2 N \sigma^2\right)}{\omega}-F^{\prime}=0
\end{equation}
\begin{equation}\label{equ9}
\frac{\omega F}{N^2 \sigma^2}+H\left(\frac{3}{r}+\frac{N^{\prime}}{N}+\frac{\sigma^{\prime}}{\sigma}\right)+H^{\prime}=0
\end{equation}
\begin{equation}\label{equ10}
\frac{8\pi H }{3 } \mu^2  r \left(\frac{F^2}{N^2 \sigma}+H^2  \sigma\right)+\sigma'\frac{dh(\psi)}{d\psi}=0
\end{equation}

\begin{equation}\label{equ11}
\frac{8\pi H r^3\left(\mu^2 F^2+N\left(H^2\left(\omega^2+\mu^2 N \sigma^2\right)-2 \omega H F^{\prime}+F'^2\right)\right)}{3 N \sigma^2}+[r^{4}h(\psi)]^{\prime}=0
\end{equation}
where $'$ denotes radial derivative and
\begin{equation}\label{equ12}
	h(\psi)\equiv\psi+\sum_{n=2}^{n_{\max}}\alpha^{n-1}\psi^n ,\quad\psi\equiv \frac{1-N(r)}{r^2} .
\end{equation}
The matter action is invariant under the global U(1) transformation 
$\mathcal{A}_{\beta}\rightarrow e^{i\alpha}\mathcal{A}_{\beta}$, where 
$\alpha$ is a constant.
By Noether’s theorem, this implies a conserved current
   \begin{equation}\label{equ3}	 
	 J^{\alpha} = \frac{i}{2}\left[\bar{\mathcal{F}}^{\alpha \beta} \mathcal{A}_
	{\beta}-\mathcal{F}^{\alpha \beta} \bar{\mathcal{A}}_{\beta}\right] \ . 
\end{equation}
The integration of the timelike component of this conserved current on a spacelike hypersurface $\Sigma$ yields a conserved quantity - the Noether charge
\begin{equation}\label{equ4}
	Q = \int_{\Sigma} J^\alpha n_\alpha \, dV \ ,
\end{equation}
where $n_\mu$ is the unit normal vector of the spacelike hypersurface $\Sigma$.
The ADM mass $M$ can be extracted from the asymptotic form of the metric components at spatial infinity. In $D=5$, the $g_{tt}$ expands as
\begin{equation}\label{equ5}
	g_{tt}=-\sigma(r)^2N(r)=-1+\frac{8GM}{3\pi r^2}+...\quad.
\end{equation}

In the absence of matter fields, the solutions of the Eqs.~(\ref{equ10}--\ref{equ11}) reduce to RBH solutions in Ref.~\cite{Bueno:2024dgm}
\begin{equation}\label{equ13}
\frac{d\sigma}{dr}=0 , \quad \quad\frac{d}{dr}\left[r^{4}h(\psi)\right]=0.
\end{equation}
Therefore, the solution has $\sigma(r)=1$, as required by the normalization of the time coordinate at spatial infinity, while the metric function $N(r)$ is determined by the algebraic equation
\begin{equation}\label{equ14}
	h(\psi)=\frac{m}{r^{4}},
\end{equation}
where $m$ is an integration constant proportional to the ADM mass of the system, given by
\begin{equation}\label{equ15}
	m=\frac{8GM}{3\pi}.
\end{equation}
Solving Eq.~\eqref{equ14}, one can obtain expressions for the metric function $N(r)$ corresponding to different correction orders. For the case $n=2$,
\begin{equation}\label{equ16}
	N_2(r)= 1-\frac{-r^2+\sqrt{\frac{32\alpha GM}{3\pi}+r^4}}{2\alpha}.
\end{equation}
For $n=3$, 
\begin{equation}\label{equ17}
\begin{split}
    N_3(r)=1-\frac{1}{6}\left(\frac{2^{2/3}\tilde{N}(r)}{\pi^{1/3}\:\alpha^{2}}
    \:-\frac{2\:r^{2}}{\alpha}\:-\:\frac{4\:(2\:\pi)^{1/3}\:r^{4}}{\tilde{N}(r)}\:\right)
\end{split} 
\end{equation}
where
\begin{equation}\label{equ18}
    \begin{split}
        \tilde{N}(r)=&\left(7\:\pi\:r^{6}\:\alpha^{3}\:+\:72G\:M\:r^{2}\:\alpha^{4}\:\right.\\
        &\left.+\:3\:\sqrt{\:r^{4}\:\alpha^{6}\:\left(9\:\pi^{2}\:r^{8}\:+\:112G\:M\:\pi\:r^{4}\:\alpha\:+\:576\:G^{2}\:M^{2}\:\alpha^{2}\:\right)}\:\right)^{1/3},
    \end{split}
\end{equation} 
In the limit $n \to \infty$,
\begin{equation}\label{equ19}
N_\infty(r)= 1-\frac{8GM r^{2}}{3\pi r^{4}+8GM\alpha}. 
\end{equation}
   \section{Boundary Conditions and Numerical Methods}\label{sec3}
        To solve the system of ordinary differential equations (ODEs) derived in Sec.~\ref {sec2}, Eqs.~(\ref{equ8}--\ref{equ11}), appropriate boundary conditions should be specified for each unknown function. We focus on static, spherically symmetric, asymptotically flat solutions, i.e., solutions whose metric approaches the Minkowski form at spatial infinity. The asymptotic behavior of the metric and Proca field at infinity is required to be
        \begin{equation}\label{equ20}
    N(\infty)=1,\quad \sigma(\infty)=1,\quad F(\infty) = 0,\quad G(\infty) = 0 .
\end{equation}

Expanding the field equations near the origin yields the boundary conditions that the metric and Proca field at the origin must satisfy    
\begin{equation}\label{equ21}
 \quad N(0)=1,\quad \sigma(0)=\sigma_0,
\end{equation}

\begin{equation}\label{equ22}
\left. \frac{dF(r)}{dr}\right|_{r = 0} = 0, 
\qquad H(0) = 0 \ . 
\end{equation}
Here, the value of $\sigma_0$ can be determined by solving the ODEs.
We introduce the following rescalings to obtain dimensionless quantities
       \begin{equation}\label{equ23}
r = r/\rho,  
\quad \omega = \omega\rho, \quad \mu = \mu\rho, 
\end{equation}
where $\rho$ is a positive constant with dimensions of length. In the numerical computations, we set $4\pi G=1$ and $\mu=1$.
To facilitate numerical computations, we compactify the radial coordinate by introducing
\begin{equation}\label{equ24}
x = \frac{r}{1+r}. 
\end{equation}
Through this transformation, the radial coordinate range $r\in[0, \infty)$ is mapped to the finite interval $x\in[0, 1]$. We employ the finite element method to numerically solve the ODEs, discretizing the integration domain $0\leq x\leq 1$ into 1000 grid points. The Newton–Raphson method is employed as the iterative method. To ensure the accuracy of the results, we impose a convergence tolerance requiring the relative error to remain below $10^{-5}$.

       \section{Numerical Results} \label{sec4}

     In this section, we show the numerical results of the model. We first consider the cases $n = 1$ and $n = 2$, which correspond to Proca stars in five-dimensional Einstein gravity~\cite{Hartmann:2010pm,
Hartmann:2012gw,
Blazquez-Salcedo:2019qrz} and Gauss–Bonnet gravity~\cite{Hartmann:2013tca,
Brihaye:2013zha,
Brihaye:2014bqa,
Brihaye:2015jja}, respectively. For higher-order curvature corrections with $n \geq 3$, we observe that the numerical solutions exhibit qualitatively similar behavior. We therefore focus on the representative cases $n=3,4,\infty$, while other higher-order cases are not shown explicitly. The following subsections discuss the spacetime geometric and physical properties of the solutions for different correction orders.

       \subsection{$n \leq 2$: Einstein gravity and Gauss–Bonnet gravity} 

       \begin{figure}[!htbp]
   	\centering
   \includegraphics[width=0.45\textwidth]{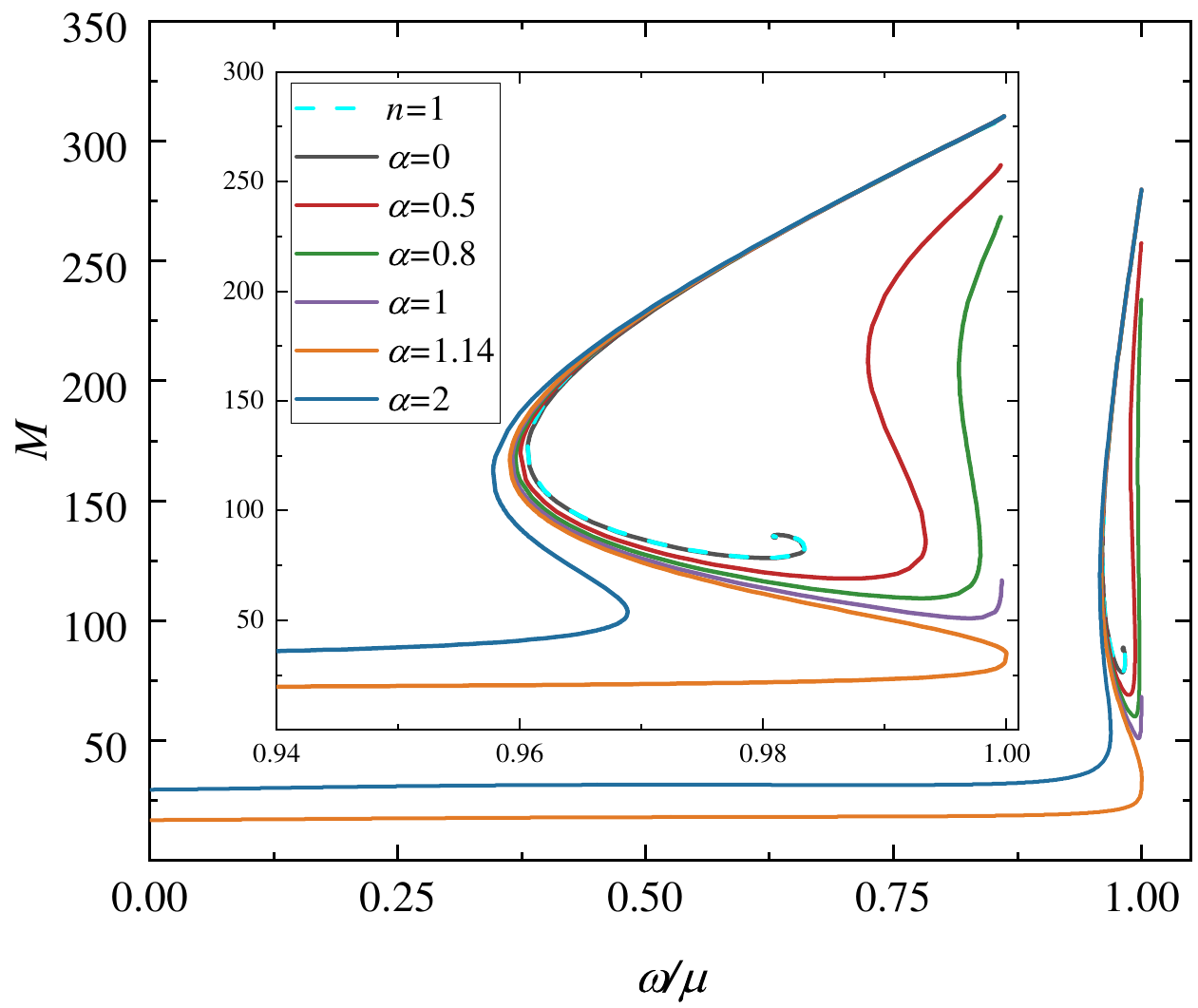}
   \includegraphics[width=0.45\textwidth]{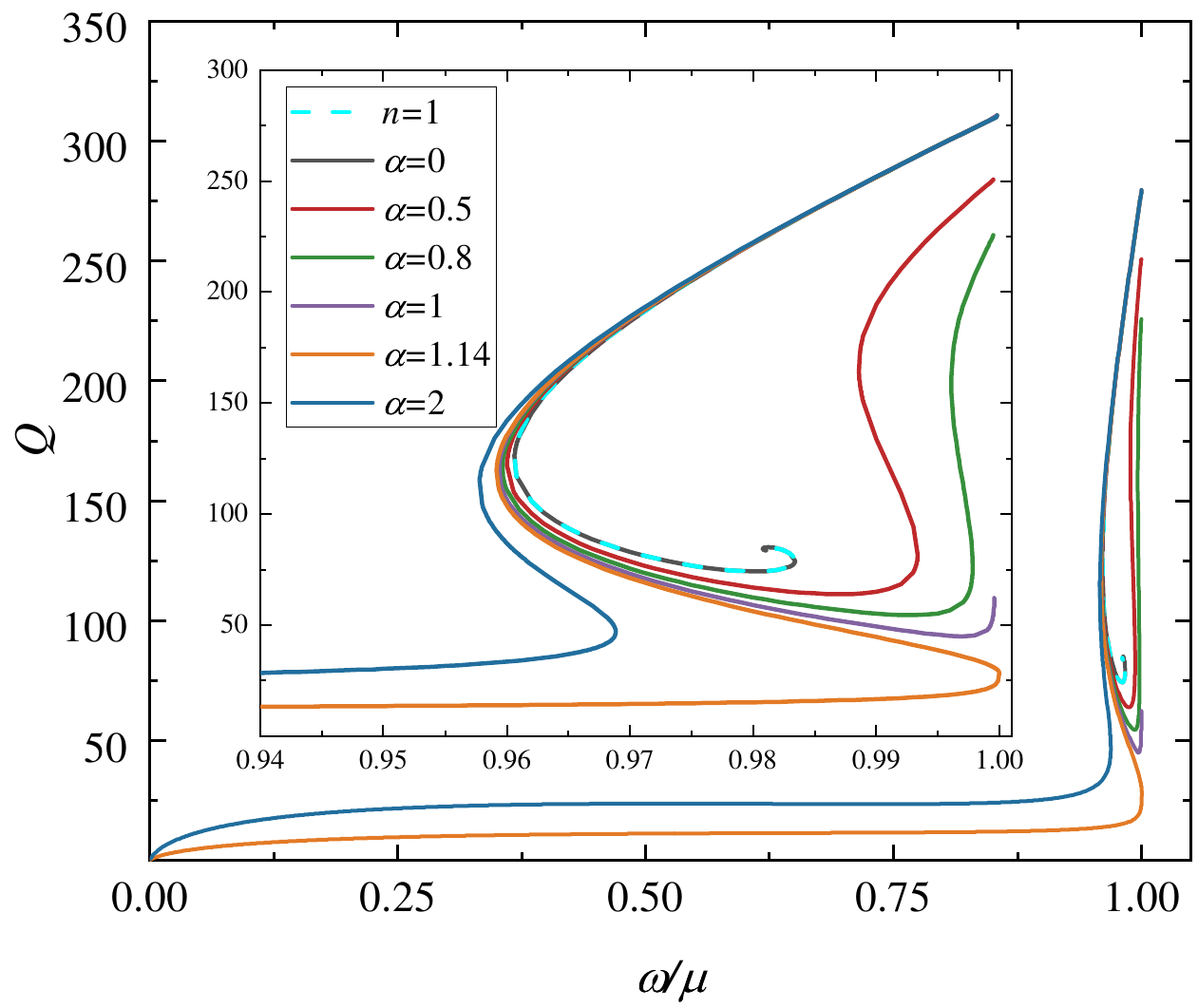}
   	\caption{The ADM mass $M$ (left panel) and Noether charge $Q$ (right panel) as functions of the frequency $\omega$ for $n=1$ (dashed) and $n=2$ (solid) with different values of the coupling $\alpha$. The insets show a magnified view of the high-frequency region near $\omega\to 1$.}
   	\label{QM_n2}
   \end{figure} 
In this section, we consider the cases for $n = 1$ and $n = 2$.
Fig.~\ref{QM_n2} displays the ADM mass (left panel) and the Noether charge (right panel) as functions of the frequency $\omega$.
The dashed curves denote the $n=1$ case, while the solid curves represent the $n=2$ solutions under different values of the coupling constant $\alpha$.
In particular, when $\alpha=0$, the higher-curvature corrections vanish and the theory reduces to five-dimensional Einstein gravity. In this limit, the $n=2$ solutions coincide with those of the $n=1$ case, providing a consistency check of the numerical results.
In five-dimensional Einstein gravity ($n=1$), the curves of the ADM mass and Noether charge versus frequency exhibit a spiral structure with multiple branches, analogous to the four‑dimensional case.
However, the limiting behavior as $\omega \to 1$ depends on the spacetime dimension.
In four dimensions, both the ADM mass and the Noether charge approach zero as $\omega \to 1$. However, in five dimensions, the Noether charge and the ADM mass are nonvanishing in the limit $\omega \to 1$, and the Noether charge approaches the ADM mass.

When the Gauss--Bonnet correction term is included ($n=2$), the spiral structure in the curves of the ADM mass and Noether charge versus frequency gradually unfolds as the coupling constant $\alpha$ increases.
For $\alpha < 1.14$, the solutions are still mainly confined to the high‑frequency region close to $\omega \to 1$.
Once $\alpha$ exceeds the critical value, i.e., $\alpha \geq 1.14$, the curves fold back at the high‑frequency and extend further toward the low-frequency regime, giving rise to solution branches that can be continued to the limit $\omega \to 0$.
With further increase of $\alpha$, the spiral structure disappears and the multi-branch behavior vanishes, and both the ADM mass and the Noether charge decrease monotonically with decreasing frequency.
For a fixed coupling constant $\alpha$, along the low-frequency branch and in the limit $\omega \to 0$, the ADM mass $M$ approaches a finite constant while the Noether charge $Q$ tends to zero.

        \begin{figure}[!]
   	\centering
     \includegraphics[width=0.45\textwidth]{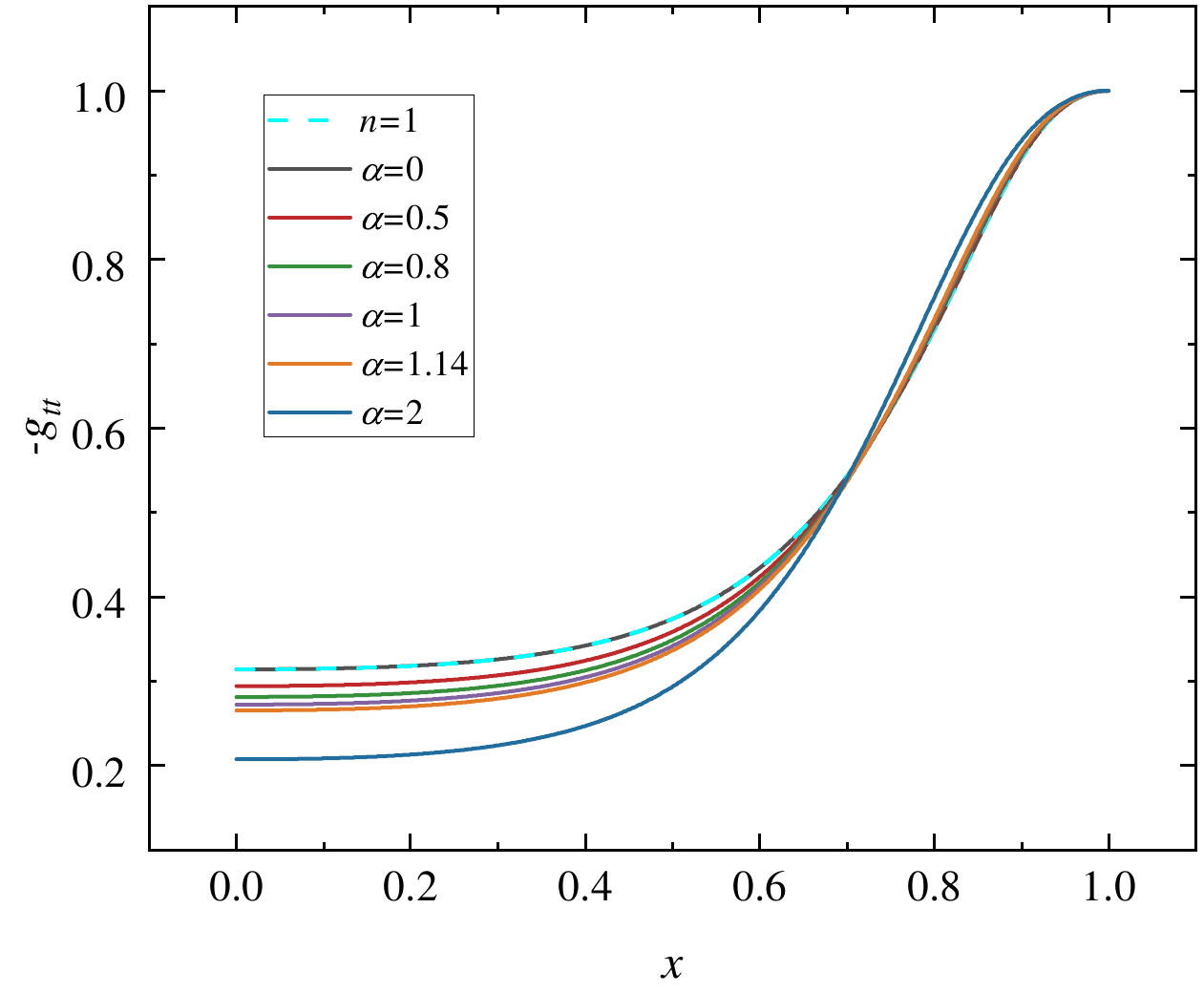}
   \includegraphics[width=0.45\textwidth]{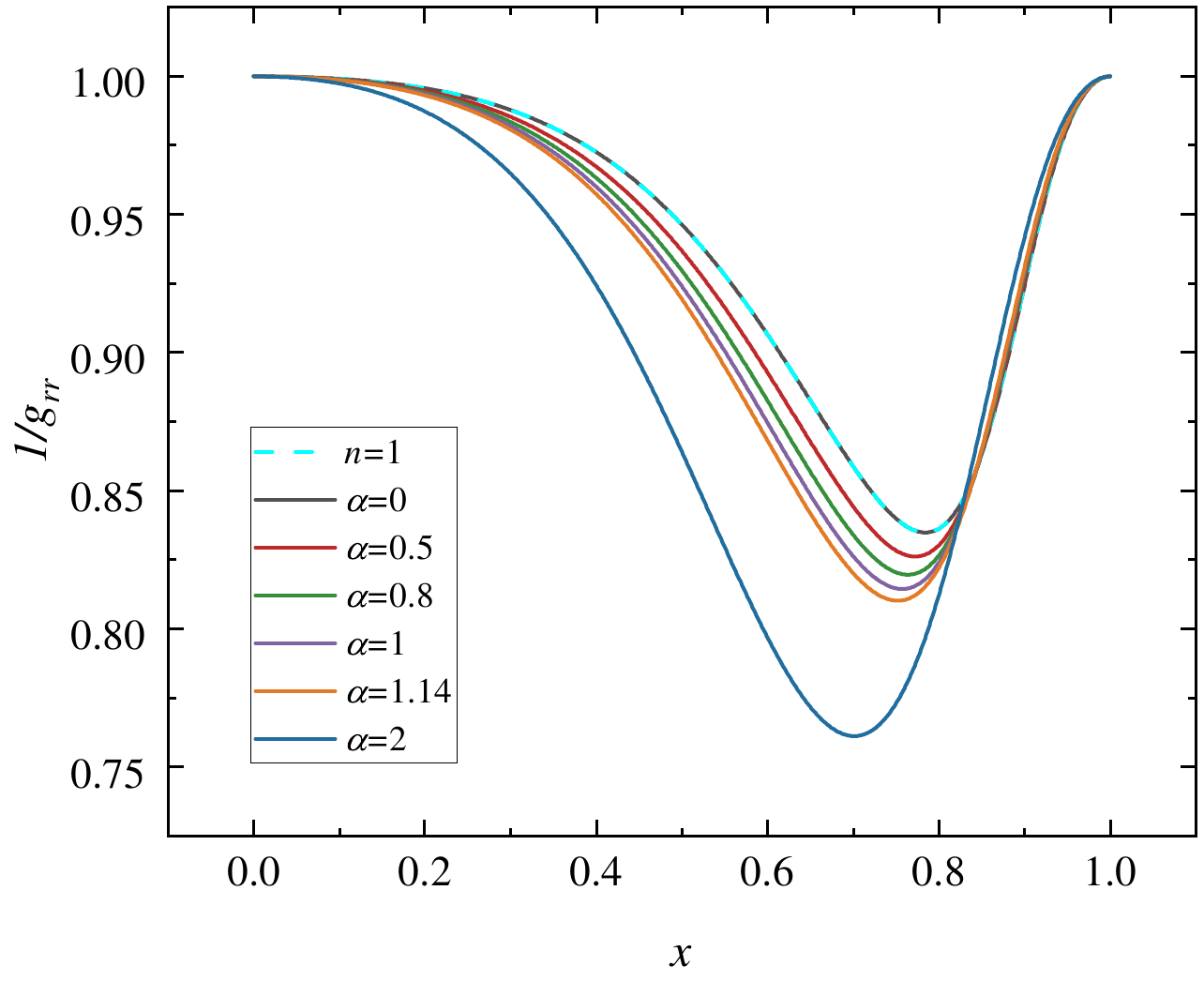}
   	\includegraphics[width=0.45\textwidth]{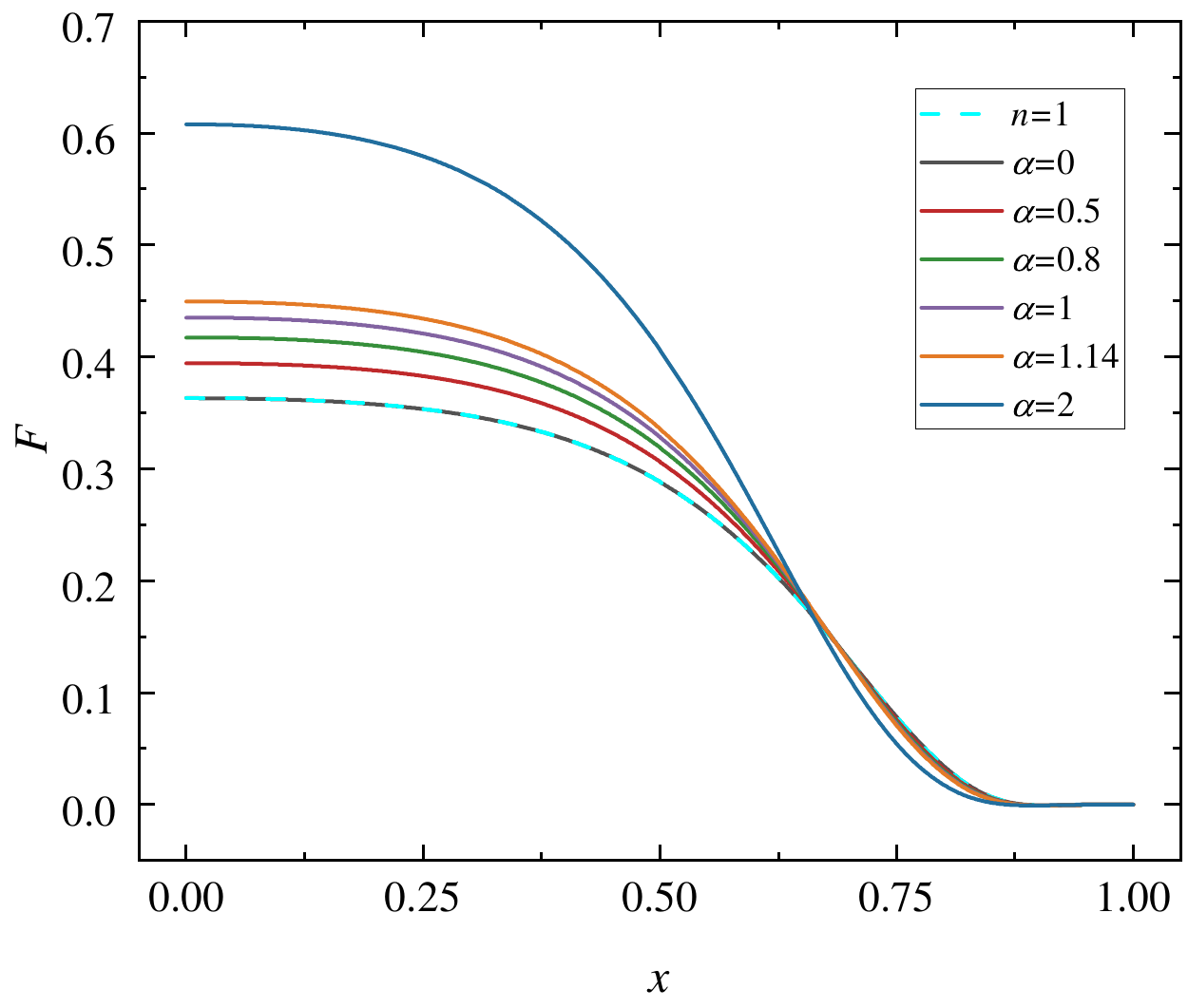}
   \includegraphics[width=0.45\textwidth]{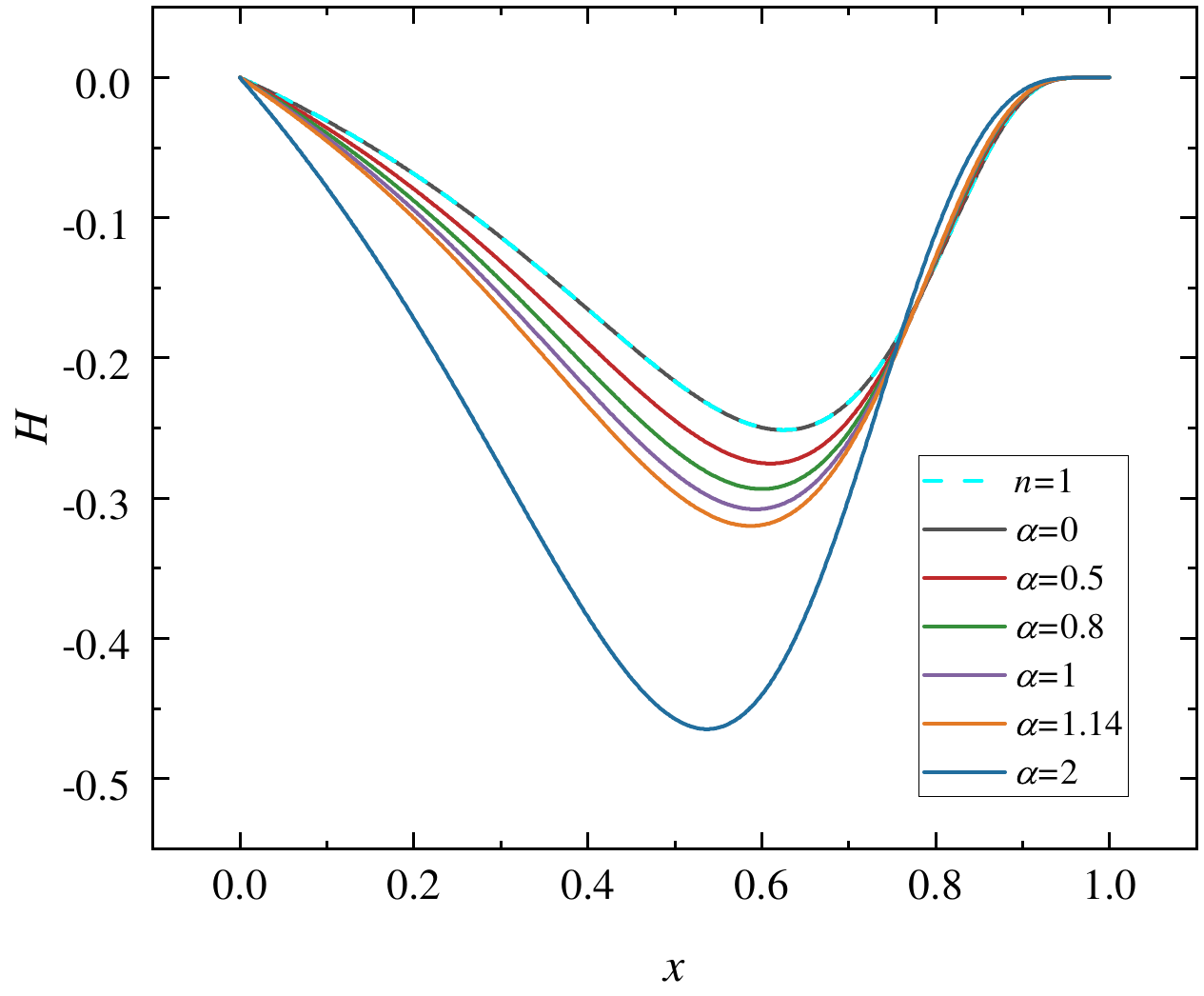}
   	\caption{  
  The metric components $-g_{tt}$ (top left) and $1/g_{rr}$ (top right), and the field functions $F(x)$ (bottom left), $H(x)$ (bottom right) as functions of the radial coordinate for $n=1$ (dashed line) and $n=2$ (solid lines) with different coupling constants $\alpha$.
All solutions correspond to the second branch with a frequency $\omega=0.965$.
}
   	\label{field_2branch}
   \end{figure} 
The global conserved quantities $M$ and $Q$ primarily characterize the overall properties of the solutions.
We now turn to investigate the influence of the Gauss–Bonnet correction term on the matter fields and the spacetime structure.
Fig.~\ref{field_2branch} displays the radial distributions of the metric components (top panels) and the Proca field (bottom panels) for $n=1$ (dashed line) and $n=2$ (solid lines) with different values of the coupling constant $\alpha$.
As shown in Fig.~\ref{QM_n2}, on the first branch, the $M$ and $Q$ curves corresponding to different $\alpha$ nearly coincide, indicating that this branch is weakly responsive to Gauss–Bonnet correction. 
Therefore, to more clearly illustrate the influence of $\alpha$ on the local structure of the solutions, we focus on the second branch, which exhibits a stronger dependence on $\alpha$, and fix the frequency to $\omega=0.965$.
At this frequency, we find that as the coupling constant $\alpha$ increases, the amplitude of the field functions grows, while the minimum values of the metric components $-g_{tt}$ and $1/g_{rr}$ decrease.
In the asymptotic region ($x \to 1$), the solutions corresponding to different values of $\alpha$ gradually coincide, indicating that they exhibit the same asymptotic behavior at spatial infinity.

        \begin{figure}[!htbp]
   	\centering
    \includegraphics[width=0.45\textwidth]{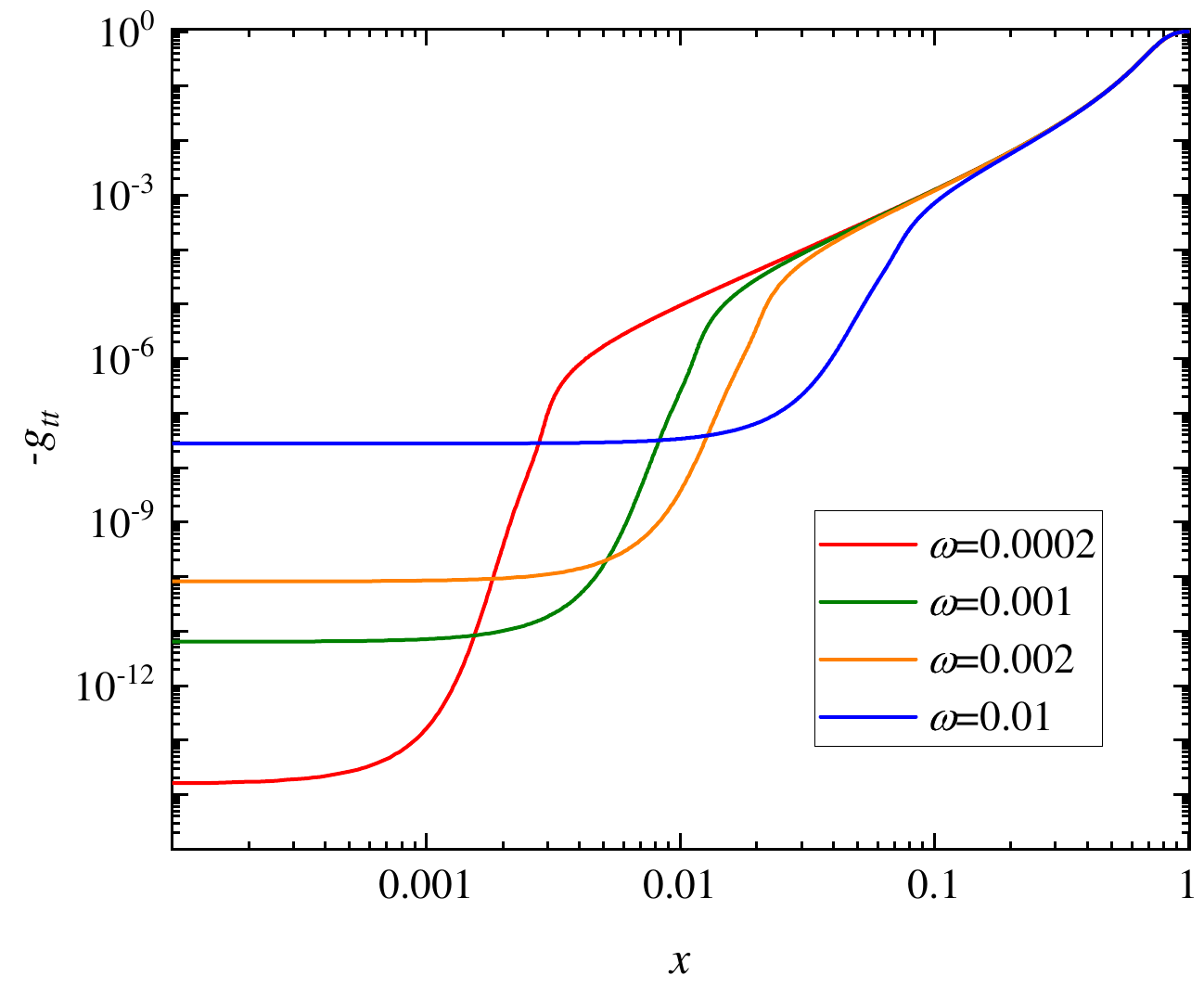}
   \includegraphics[width=0.45\textwidth]{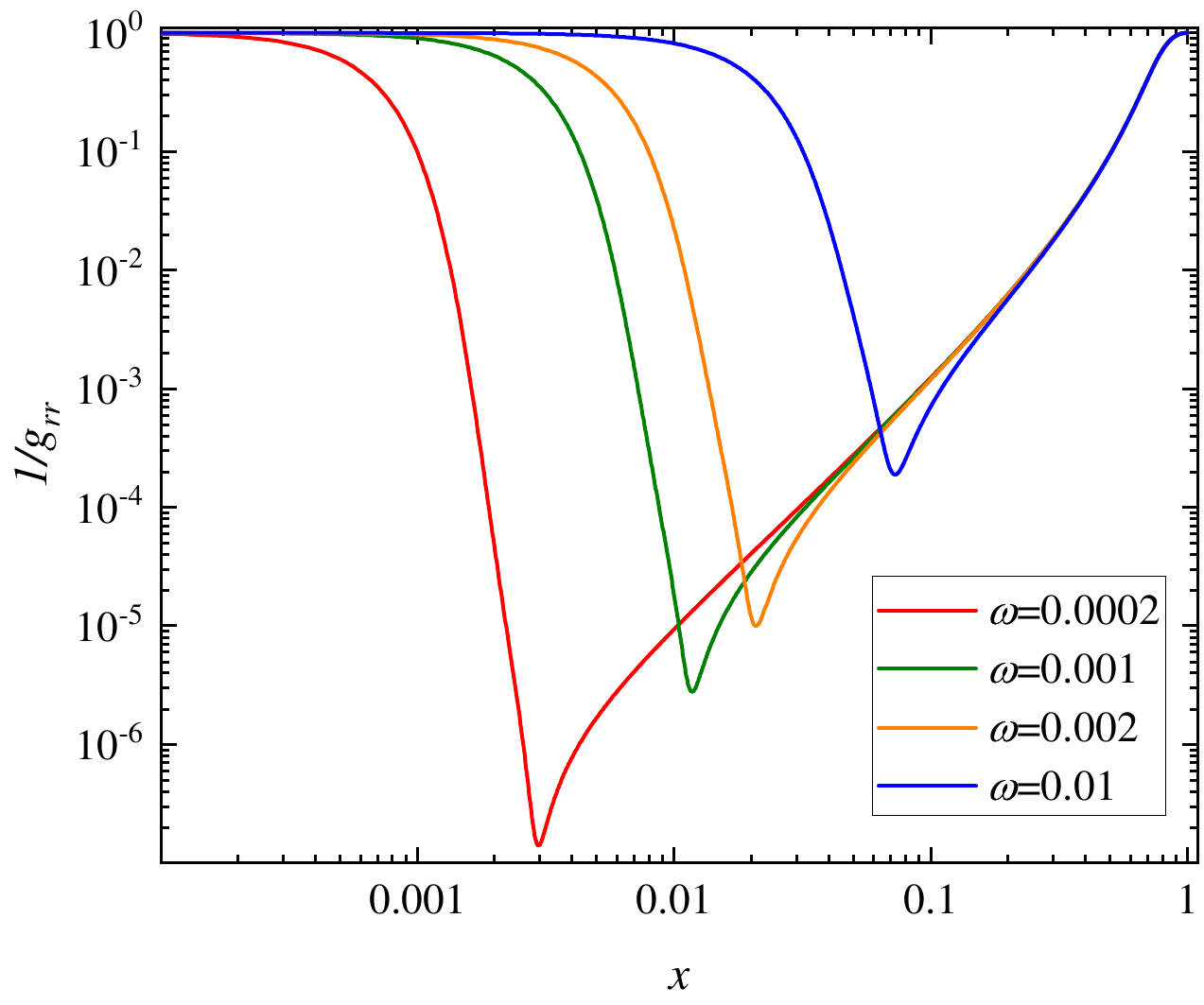}
   	\includegraphics[width=0.45\textwidth]{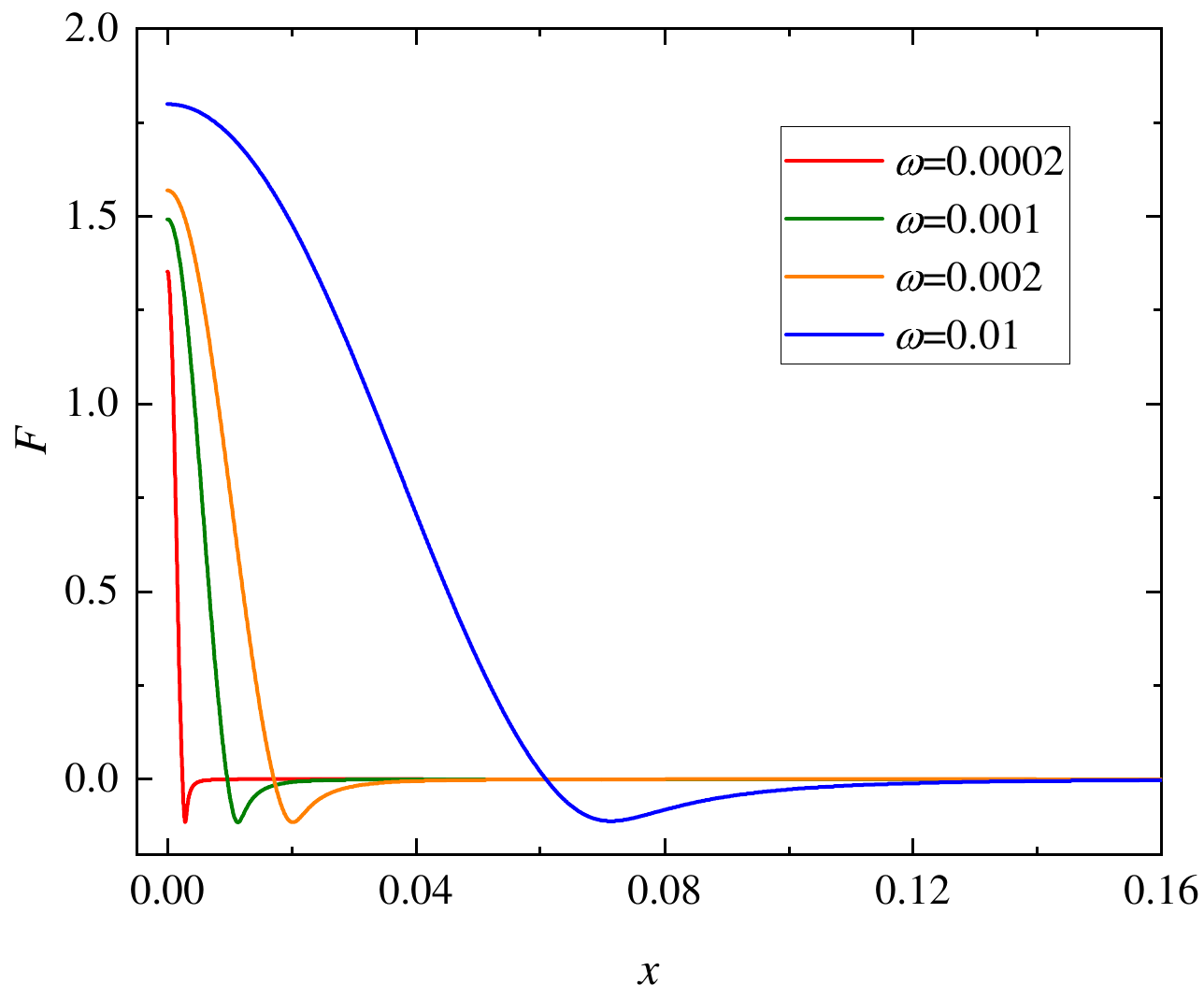}
   \includegraphics[width=0.45\textwidth]{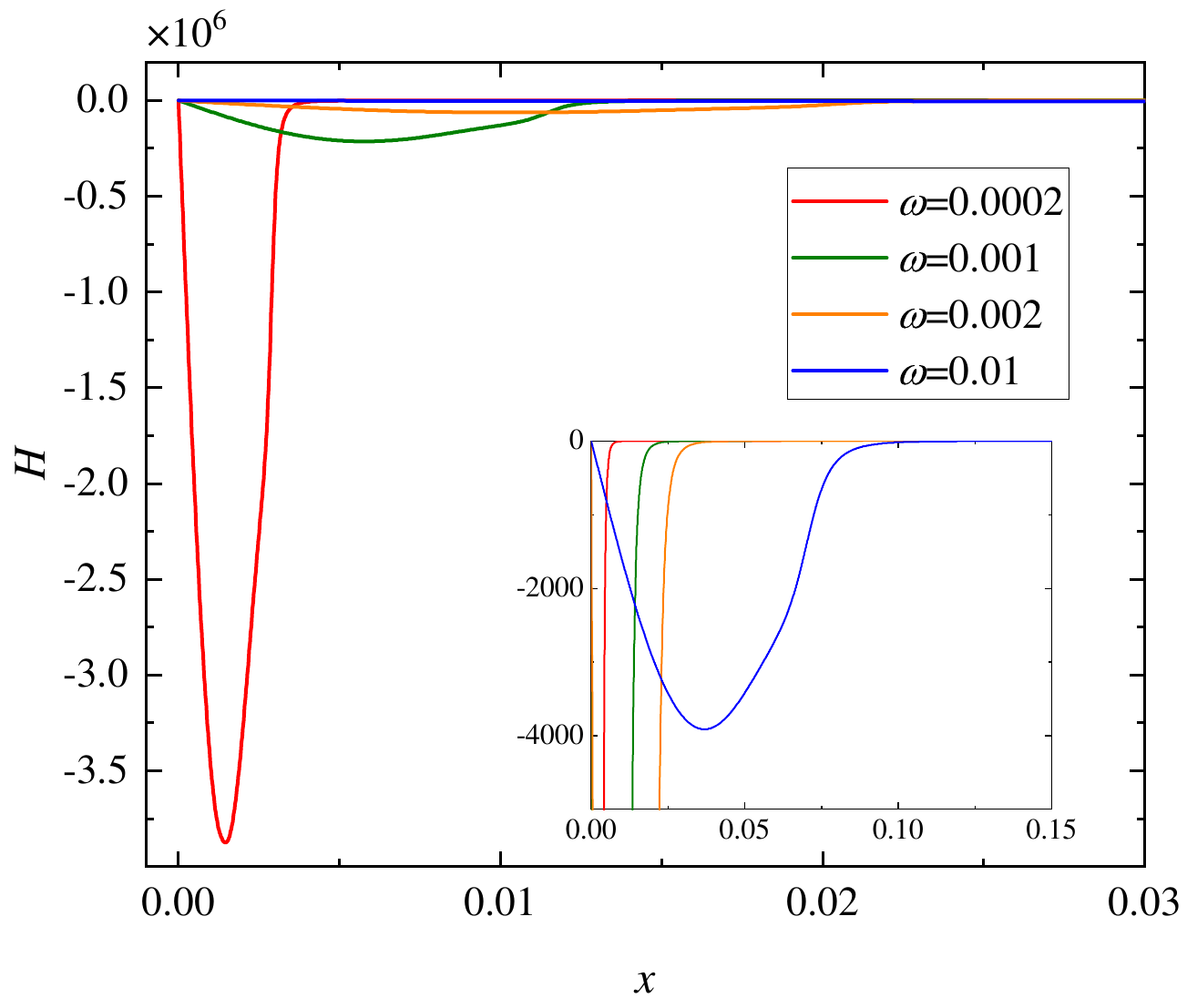}
   	\caption{The metric components $-g_{tt}$ (top left) and $1/g_{rr}$ (top right), and the field functions $F(x)$ (bottom left), $H(x)$ (bottom right) as functions of the radial coordinate for $n=2$ with different frequencies $\omega$. All solutions have $\alpha=5$.}
   	\label{function_n2}
   \end{figure} 
To better illustrate the behavior of the field configurations and the spacetime geometry in the low-frequency regime, 
Fig.~\ref{function_n2} presents the radial distributions of the metric components (top panels) and the field functions (bottom panels) for the $n=2$ case at frequencies $\omega=\{0.01, 0.002, 0.001, 0.0002\}$ with $\alpha=5$. 
The Proca field is mainly localized within a finite radial region, and its amplitude decreases as the frequency increases. 
In the $\omega \to 0$ limit, the field functions of the Gauss--Bonnet Proca stars exhibit a clear divergent behavior. At the same time, the metric functions undergo significant changes. 
In particular, $1/g_{rr}$ displays an extremely steep variation near the central region, 
where its value drops rapidly within a very small radial scale, forming a nearly sharp structure. 
This indicates that the second-order curvature correction is insufficient to suppress the divergence of the matter field and leads to a sharp variation of the metric functions $1/g_{rr}$ near the center.
   
\subsection{$n \geq 3$: higher-curvature corrections}

\begin{figure}[!htbp]
   	\centering
   	\includegraphics[width=0.45\textwidth]{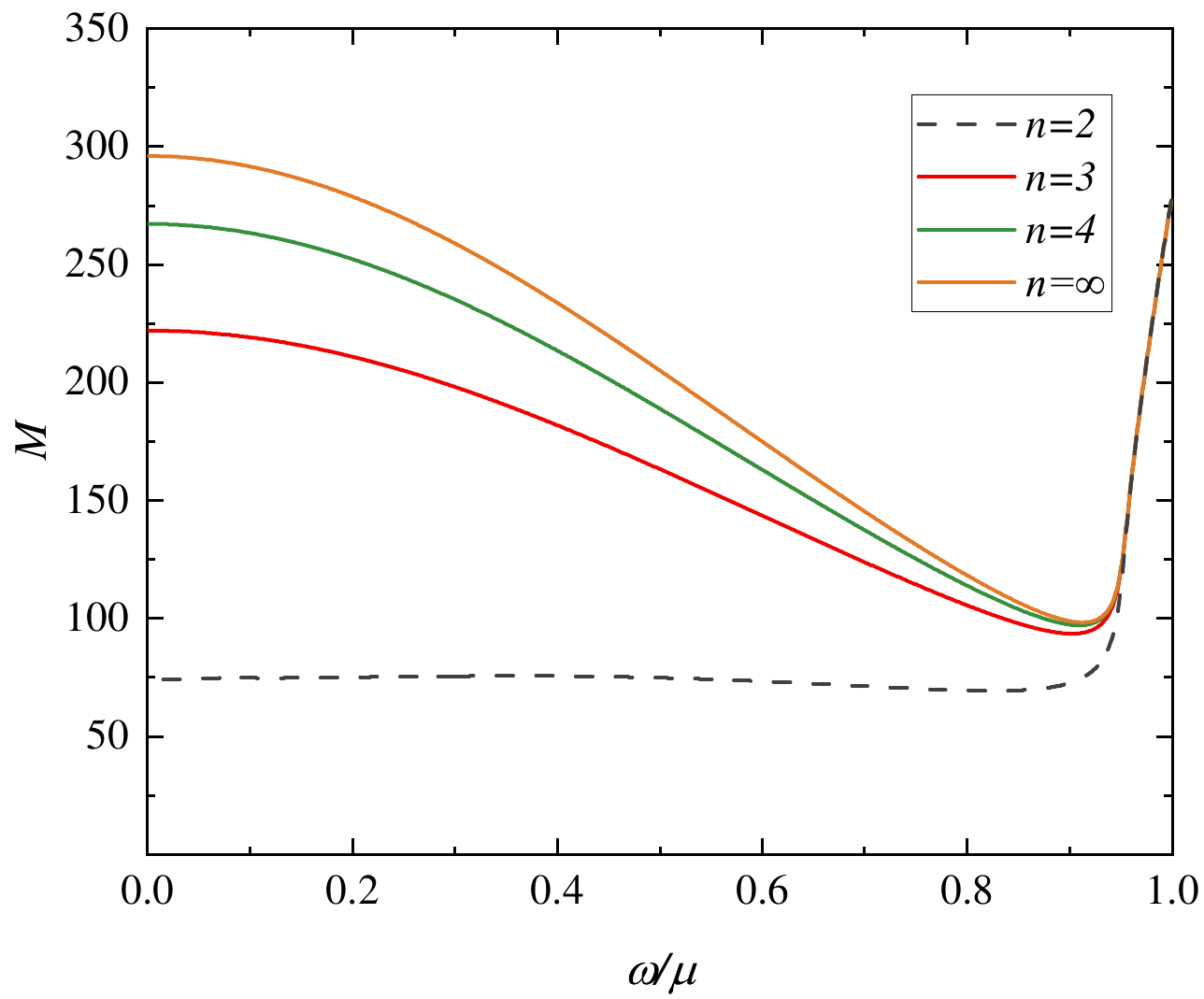}
    \includegraphics[width=0.45\textwidth]{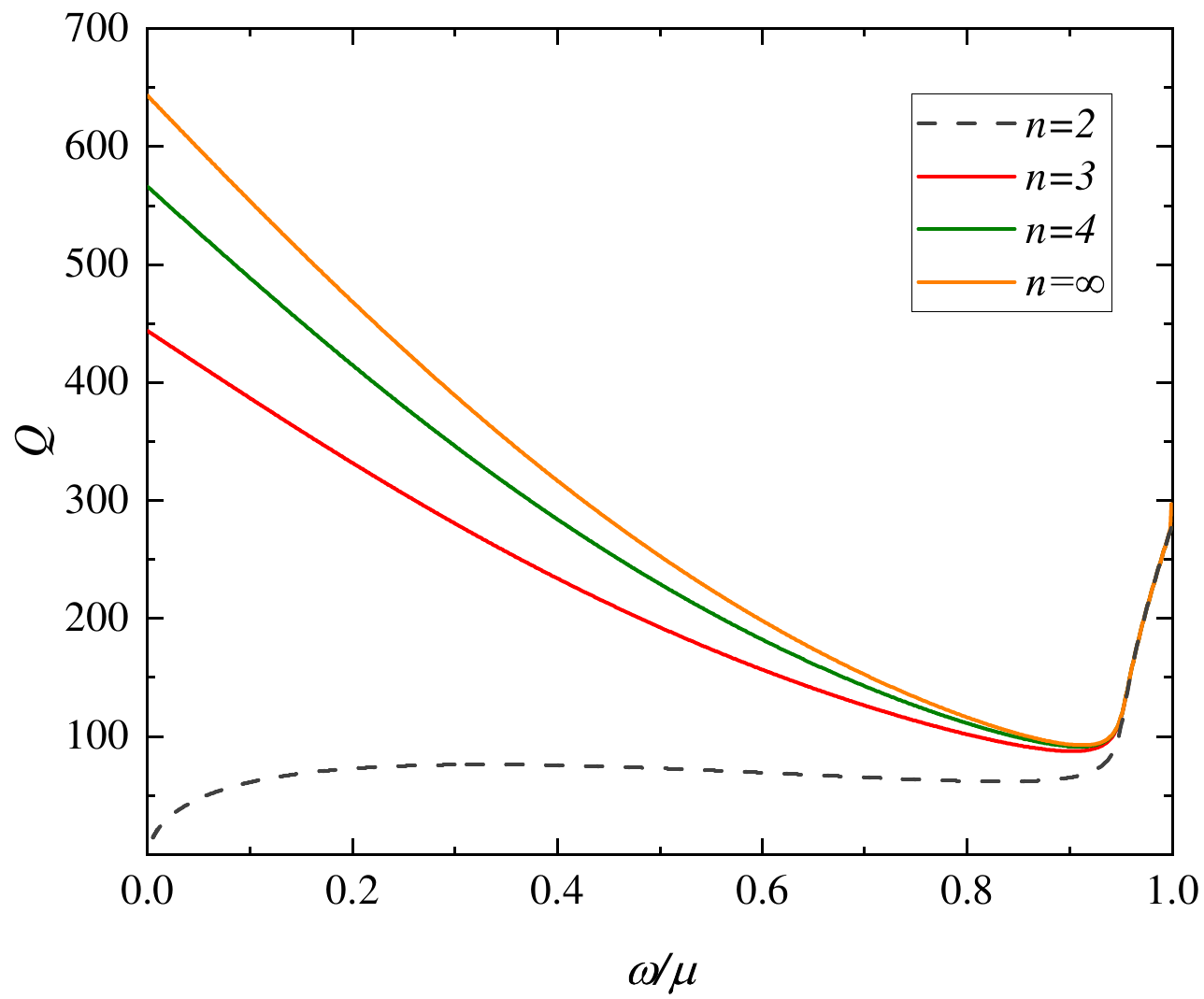}
   	\caption{The ADM mass $M$ (left panel) and Noether charge $Q$ (right panel) as functions of the frequency $\omega$ for $n=2,3,4,\infty$. Dashed curves denote $n=2$, while solid curves correspond to higher-order corrections. All solutions have $\alpha=5$.}
   	\label{MQ}
   \end{figure}

In this section, we investigate the properties of the solutions for $n=3,4,$ and $\infty$. 
We first analyze the influence of the correction order $n$ on the global conserved quantities. 
Fig.~\ref{MQ} shows the dependence of the ADM mass $M$ and the Noether charge $Q$ on the frequency $\omega$ 
for different $n$ with $\alpha=5$. 
The dashed curves correspond to the $n=2$ case. As discussed in the previous subsection, for the five-dimensional Einstein gravity case ($n=1$), 
the relation between $Q$ and $M$ exhibits a spiral structure in the frequency interval $[0.96,1)$. 
For the $n=2$ case, when the coupling constant $\alpha$ exceeds a critical threshold, the frequency domain extends to $(0,1)$ and the spiral structure gradually disappears. 
For the cases with $n\geq3$ at $\alpha=5$, both $M$ and $Q$ display a nonmonotonic variation,
first decreasing and then increasing as $\omega$ grows. In the limit $\omega \rightarrow 0$, the values of both $Q$ and $M$ increase with the correction order $n$.

     \begin{figure}[!htbp]
   	\centering
   \includegraphics[width=0.45\textwidth]{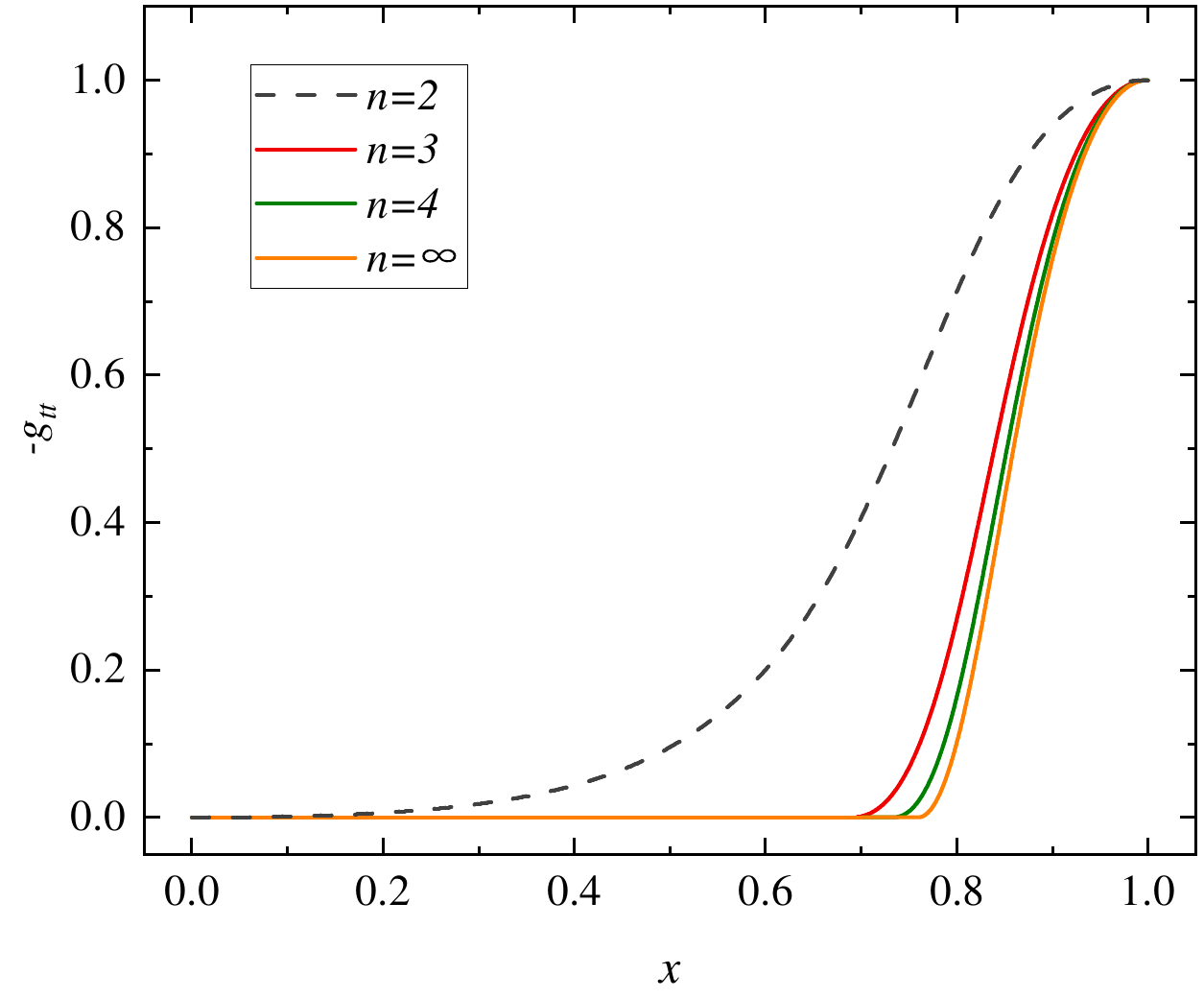}
   \includegraphics[width=0.45\textwidth]{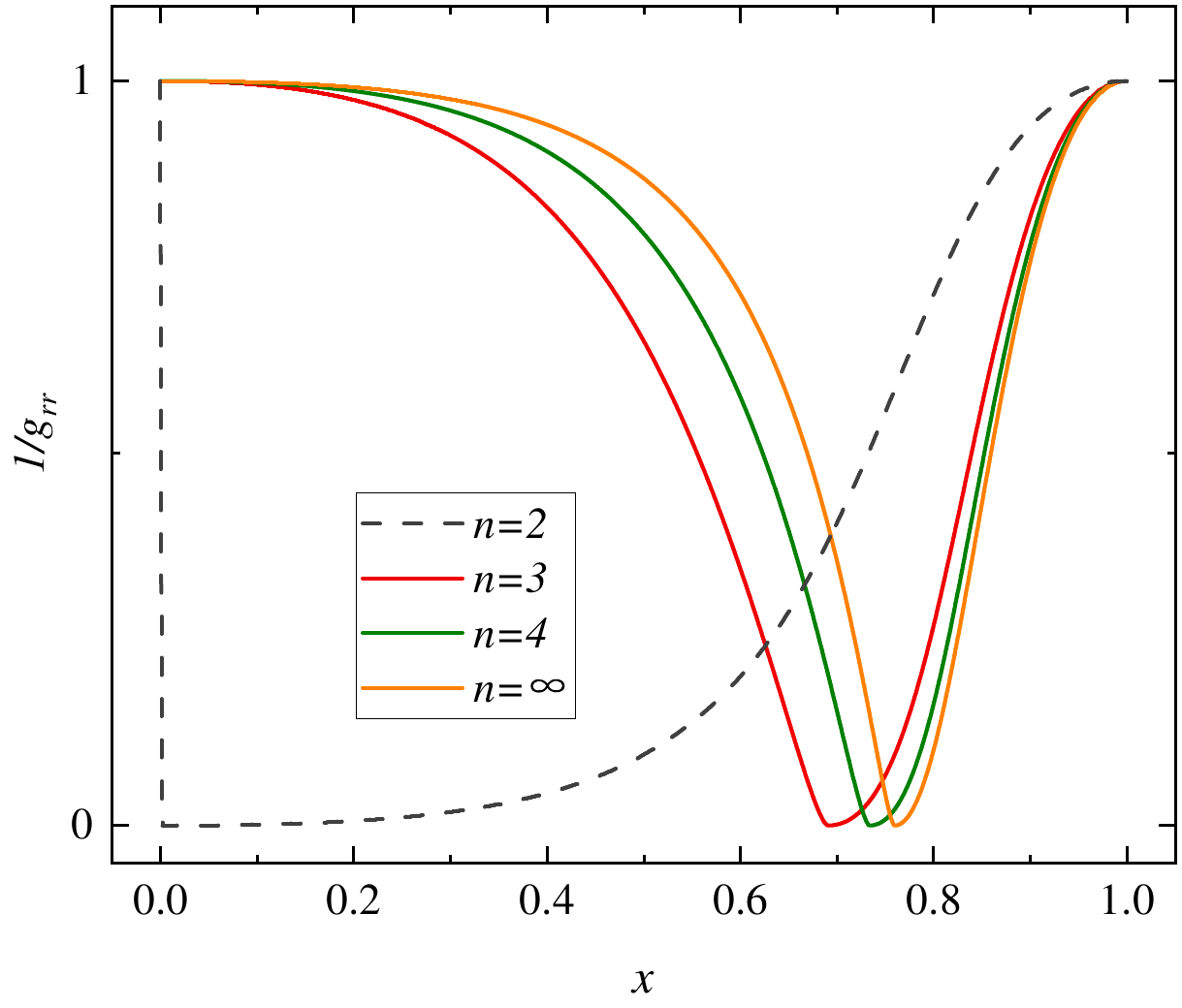}
    	\includegraphics[width=0.45\textwidth]
   {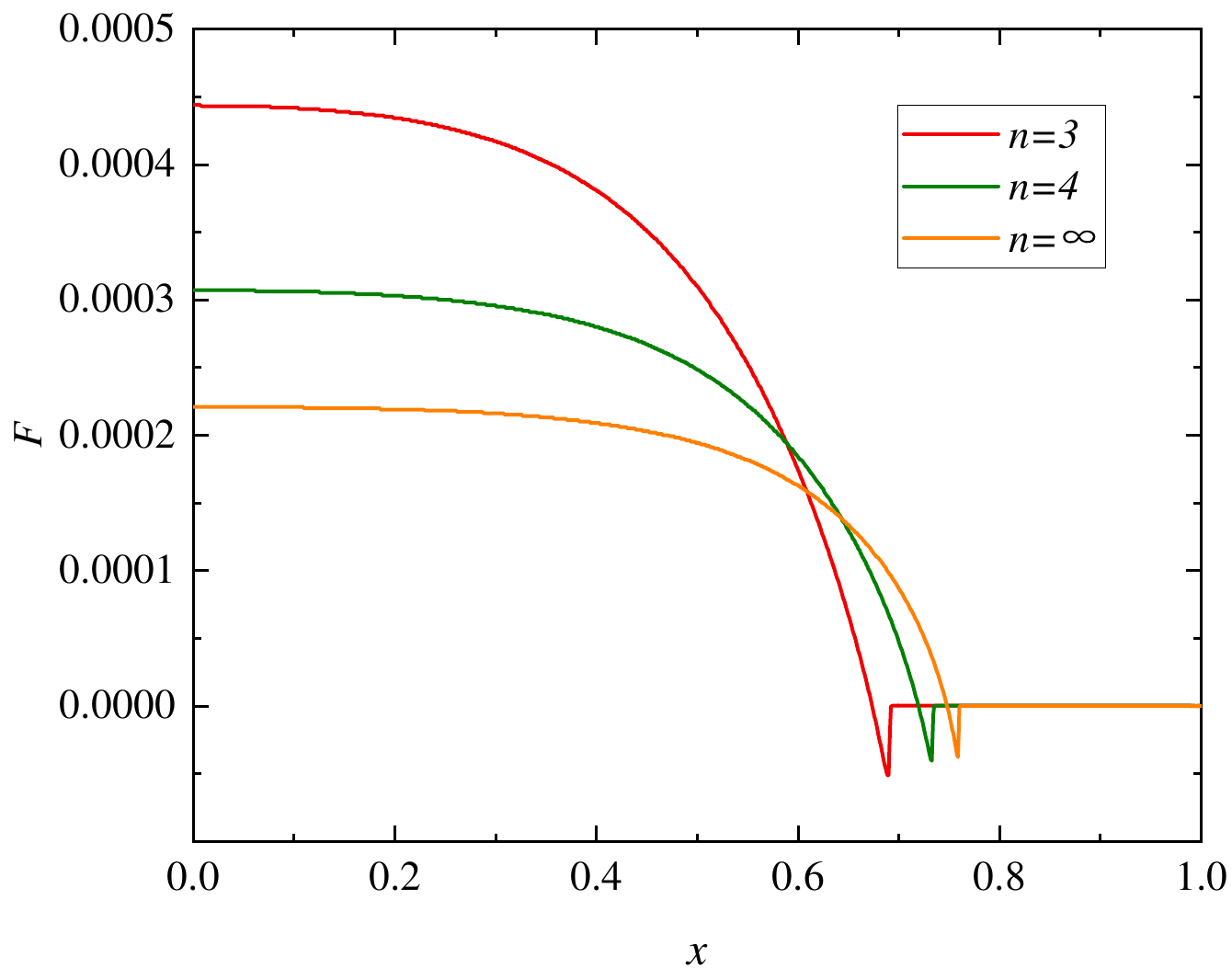}
   \includegraphics[width=0.45\textwidth]{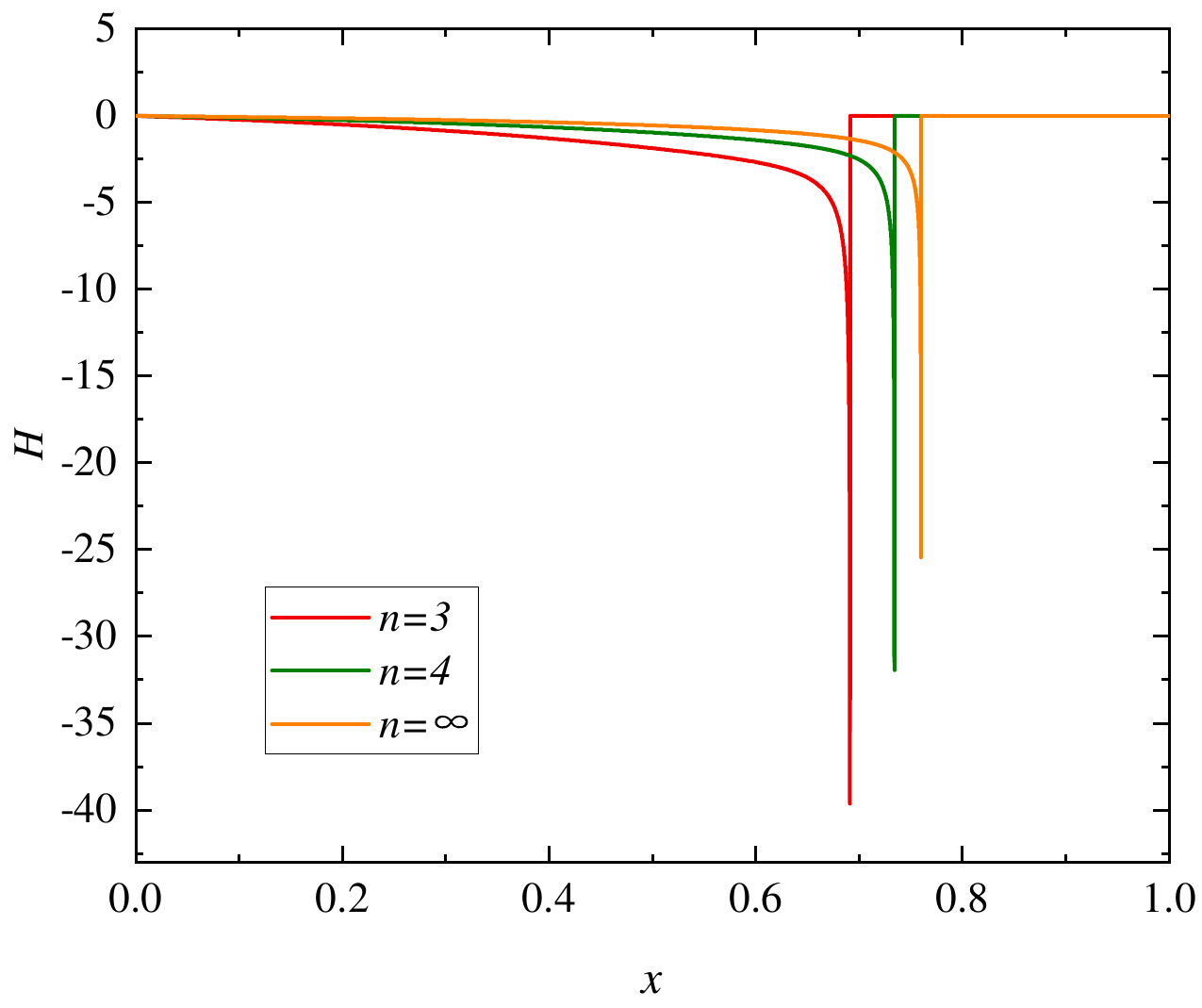}
   	\caption{The metric components $-g_{tt}$ (top left) and $1/g_{rr}$ (top right), and the field functions $F(x)$ (bottom left), $H(x)$ (bottom right) as functions of the radial coordinate for $n=3,4,\infty$. Dashed curves denote $n=2$. All solutions have $\omega=0.0002$ and $\alpha=5$.}
   	\label{field_frozen}
   \end{figure} 

Next, we investigate the field distribution and spacetime geometric features under the higher-order curvature corrections in $\omega\to 0$. Fig.~\ref{field_frozen} shows the metric (top panels) and the field distribution (bottom panels) as functions of the radial coordinate $x$ for different correction orders $n$. All solutions are set with a frequency $\omega=0.0002$ and a coupling constant $\alpha=5$ to compare the. For comparison, we show the $n=2$ case with dashed curves. We can find that for $n \ge 3$, the higher-curvature terms significantly modify the spacetime geometric behavior near the central region. 
Both the temporal component $-g_{tt}$ and the inverse of the spatial component $1/g_{rr}$ remain finite and smooth across the radial domain.
At the critical radius $r_c$, $-g_{tt}$ and $1/g_{rr}$ both tend toward zero but do not strictly reach zero, forming a critical horizon. 
Inside the critical horizon, as $-g_{tt}$ approaches zero, the gravitational redshift becomes extremely strong. To a distant observer, dynamical processes near the critical horizon are drastically slowed down, and the star appears frozen. For this reason, the solutions with frequency approaching zero are referred to as frozen stars.
The critical horizon radius $r_c$ shifts outward with increasing $n$.

Moreover, in the limit $\omega \to 0$, the behavior of the matter field differs from the $n=2$ case, where divergences are observed. 
For $n \geq 3$, the matter field remains finite over the entire radial domain. 
As the correction order $n$ increases, the amplitude of the matter field in the central region is suppressed, while the location of the critical horizon shifts outward. 
Notably, the matter field is primarily localized inside the critical horizon. Its distribution gradually decays with increasing radial distance, drops sharply at the critical horizon, and approaches zero outside it. This spatial distribution, which decays sharply at the critical horizon, is consistent with the behavior of the metric components. The distribution of metric and matter field together illustrates the impact of higher-order curvature corrections on spacetime geometry.
As the source term in the Einstein equations, the matter field distribution governs the spacetime curvature, while its own distribution is influenced by the higher-order curvature terms.
    \begin{figure}[!htbp]
   	\centering	\includegraphics[width=0.45\textwidth]{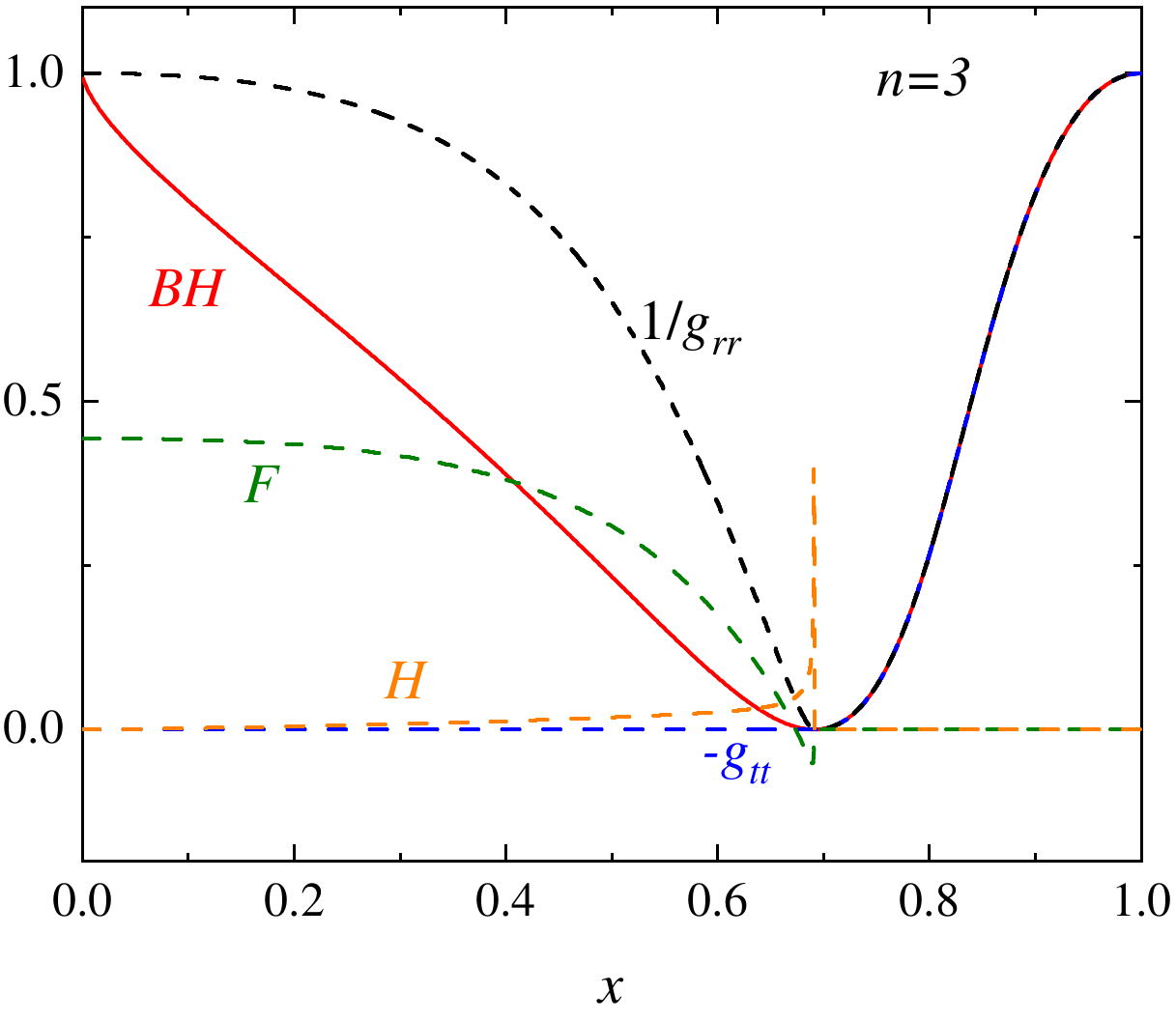}
    \includegraphics[width=0.45\textwidth]{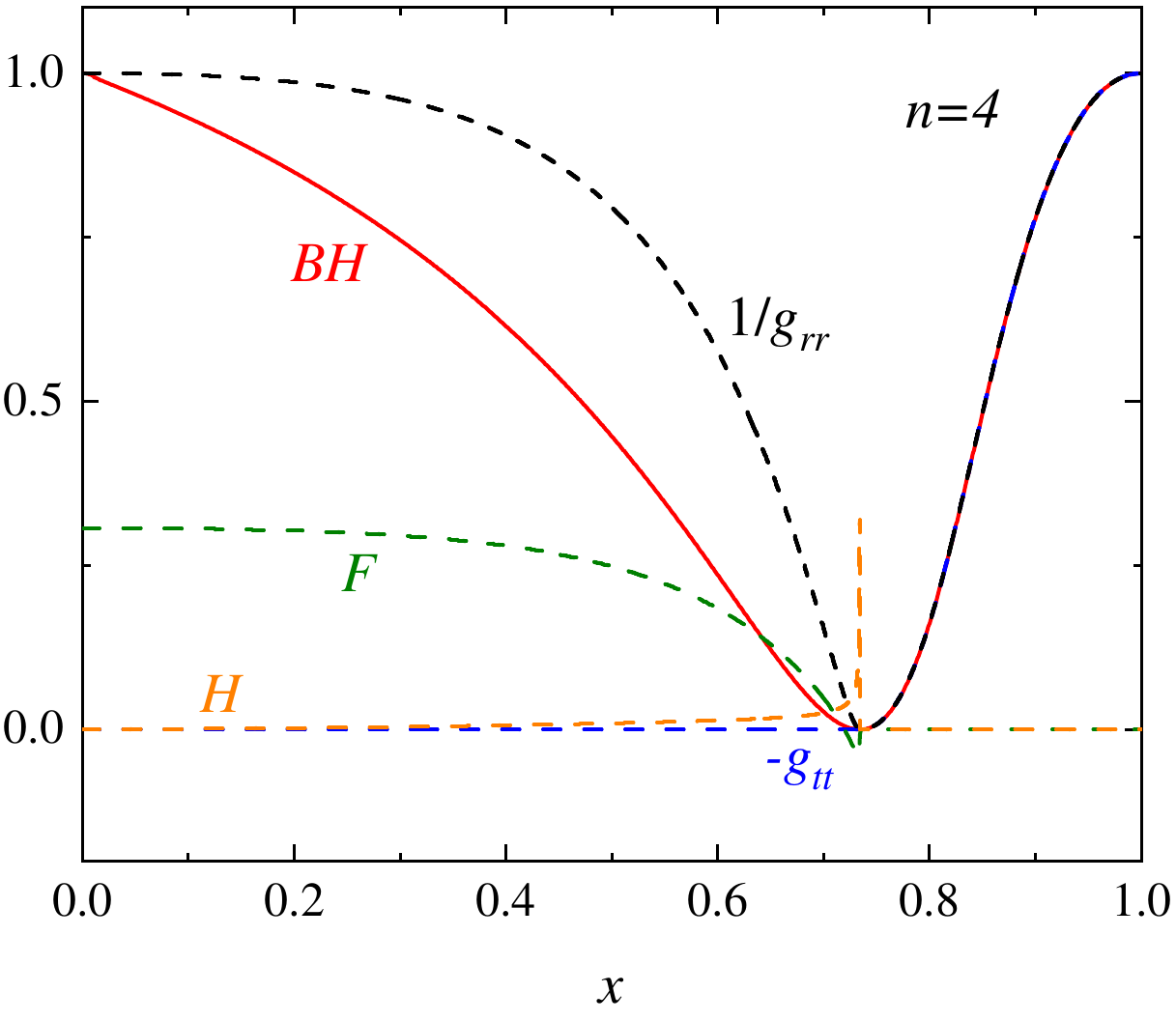}
    \includegraphics[width=0.45\textwidth]{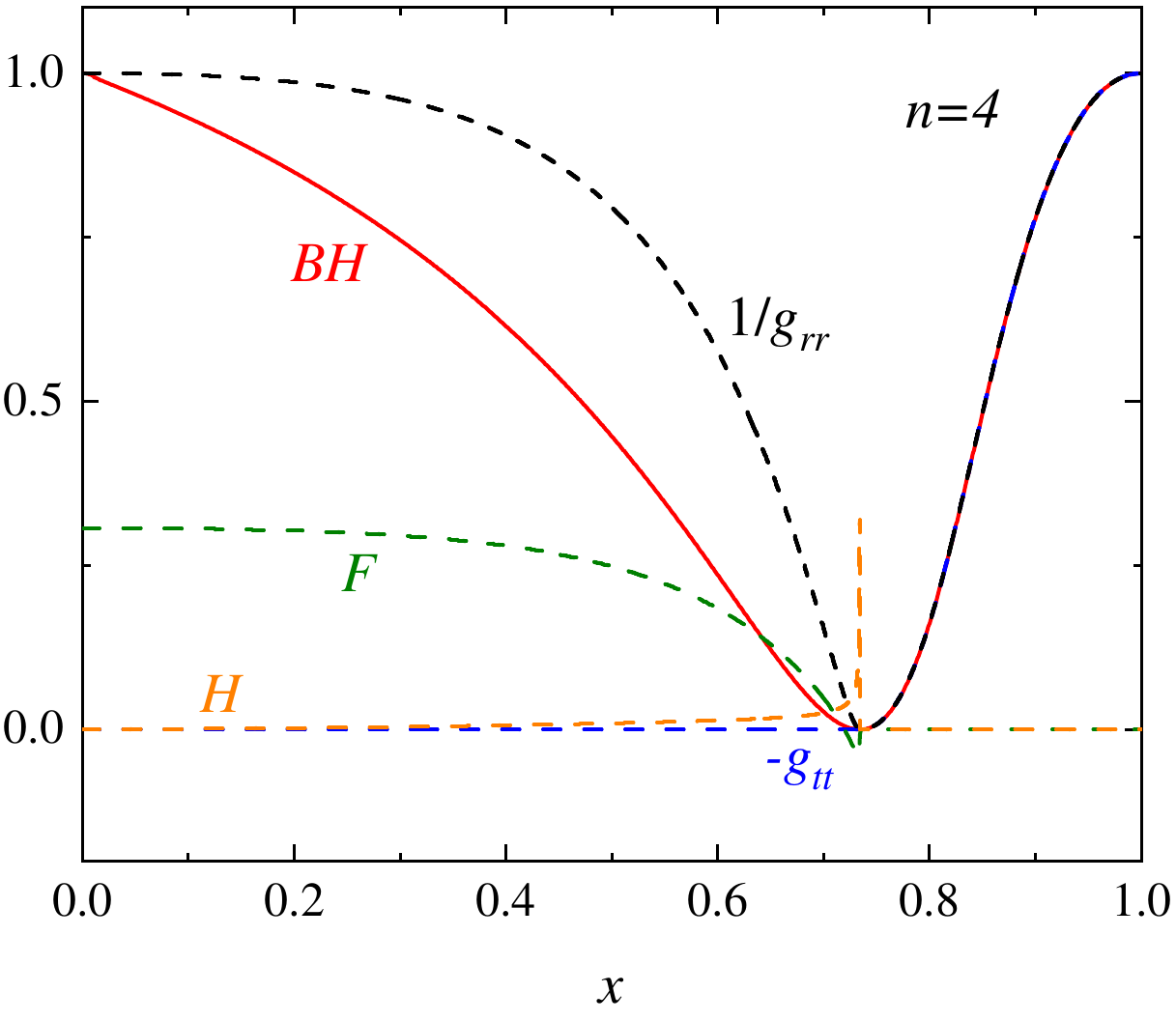}
   	\caption{ Comparison between frozen star solutions and the black hole solutions. 
The solutions are shown for $n=3$ (top left panel), $n=4$ (top right panel), and $n=\infty$ (bottom panel). 
The dashed curves denote the frozen star solutions, while the solid curves correspond to extremal black hole solutions with the same ADM mass and coupling constant $\alpha$. 
All curves are shown for $\omega = 0.0002$ and $\alpha = 5$.}
   	\label{distribution}
   \end{figure}   
   
For $n \geq 3$, frozen star solutions can already be obtained in finite-order higher-curvature corrections, without the necessity of including infinite-order curvature terms. To better understand the properties of frozen star solutions, we compare them in Fig.~\ref{distribution} with black hole solutions having the same ADM mass and coupling constant $\alpha$, where the black hole metric is given by Eq.~(\ref{equ16}), Eq.~(\ref{equ17}), and Eq.~(\ref{equ19}). For this parameter choice, the corresponding black hole solution is an extremal black hole. For clarity of presentation, the matter field function has been appropriately rescaled in the figure: the displayed $H(x)$ curve has been magnified by a factor of $10^{3}$, while $G(x)$ is shown as $-0.1$ times its numerical solution. This rescaling is introduced solely for visual purposes and does not affect the physical properties of the solutions. This figure shows that outside the critical horizon, the metric functions $-g_{tt}$ and $1/g_{rr}$ of the frozen star coincide with those of the extremal black holes, indicating that their exterior spacetime geometry is identical.  Therefore, any classical observation relying on the external spacetime geometry cannot distinguish the frozen star from the extremal black hole. 
Despite this indistinguishability in their exterior spacetime geometry, frozen stars differ fundamentally from black holes in their interior properties. In particular, frozen stars are compact objects without a curvature singularity or event horizon. As a result, they are regular everywhere and avoid the information loss paradox associated with the existence of an event horizon.

       \subsection{Energy Density and Compactness}

   \begin{figure}[!htbp]
   	\centering
    \includegraphics[width=0.45\textwidth]{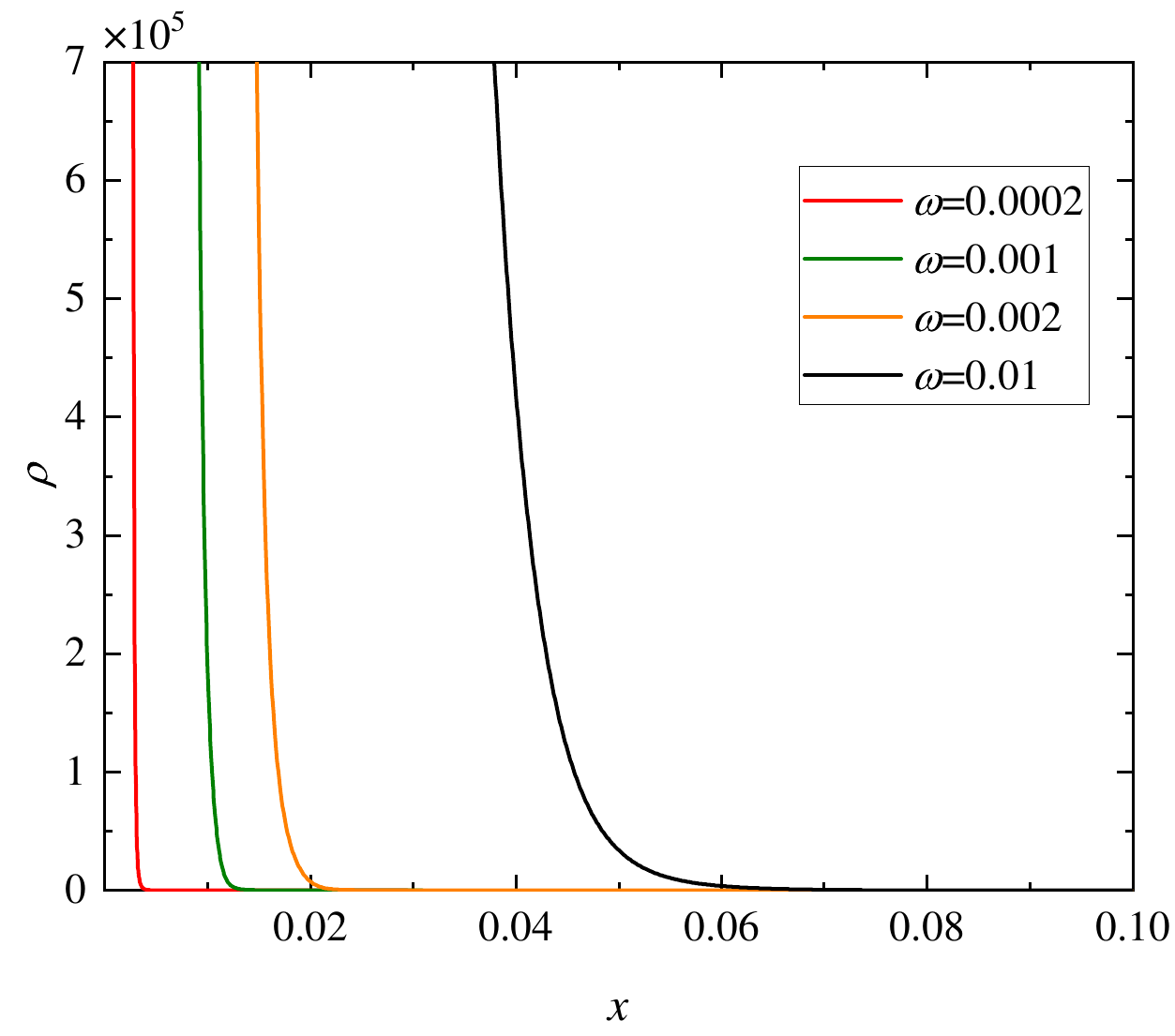}
  \includegraphics[width=0.45\textwidth]{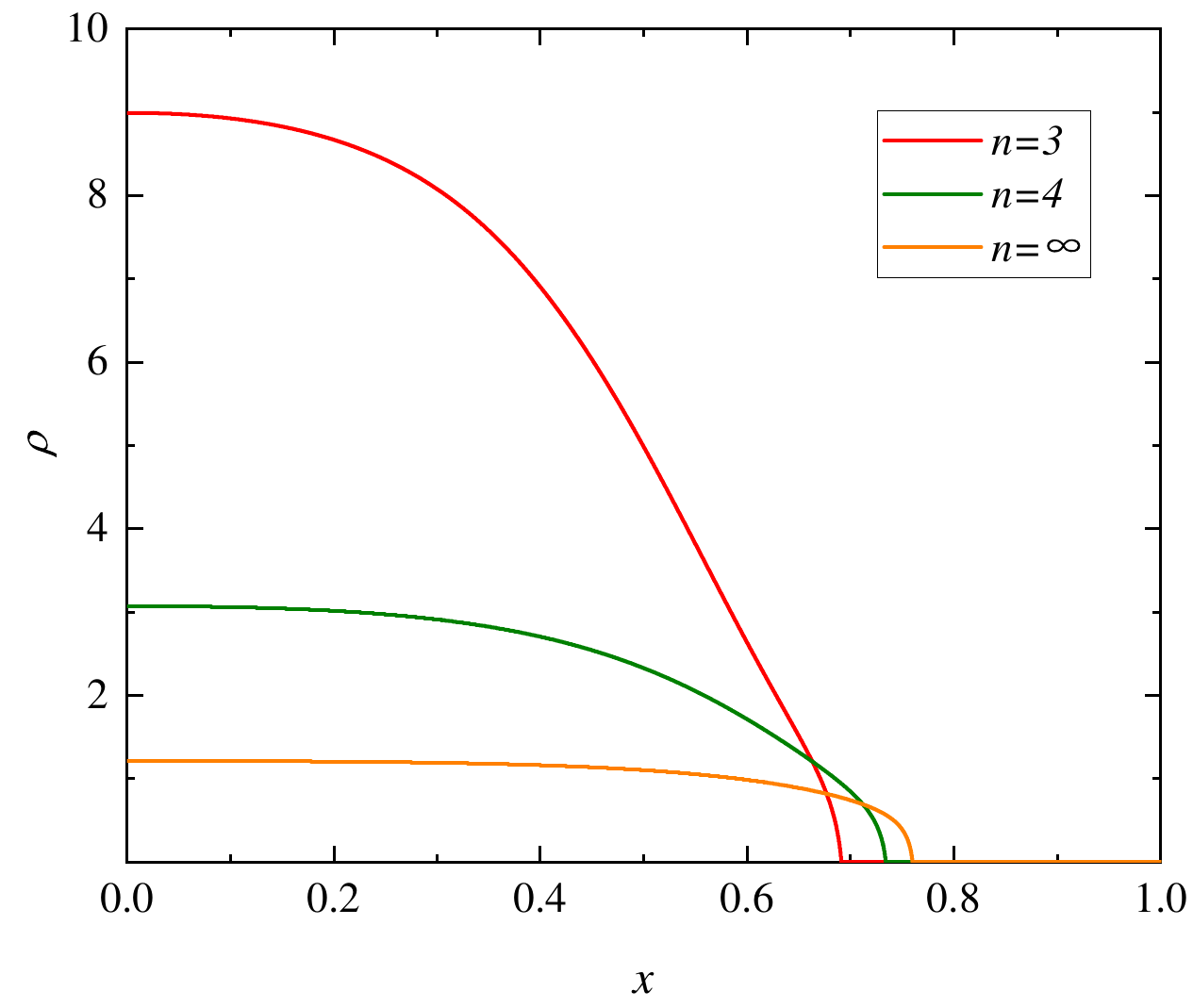}
   	\caption{The radial distribution of the energy density. Left panel: Energy density for different frequencies with $n=2$. Right panel: Energy density for correction orders $n=3,4,\infty$ at $\alpha=5$ and $\omega=0.0002$.}
   	\label{Edensity}
   \end{figure} 

\begin{figure}[!h]
    \centering	\includegraphics[width=0.45\textwidth]{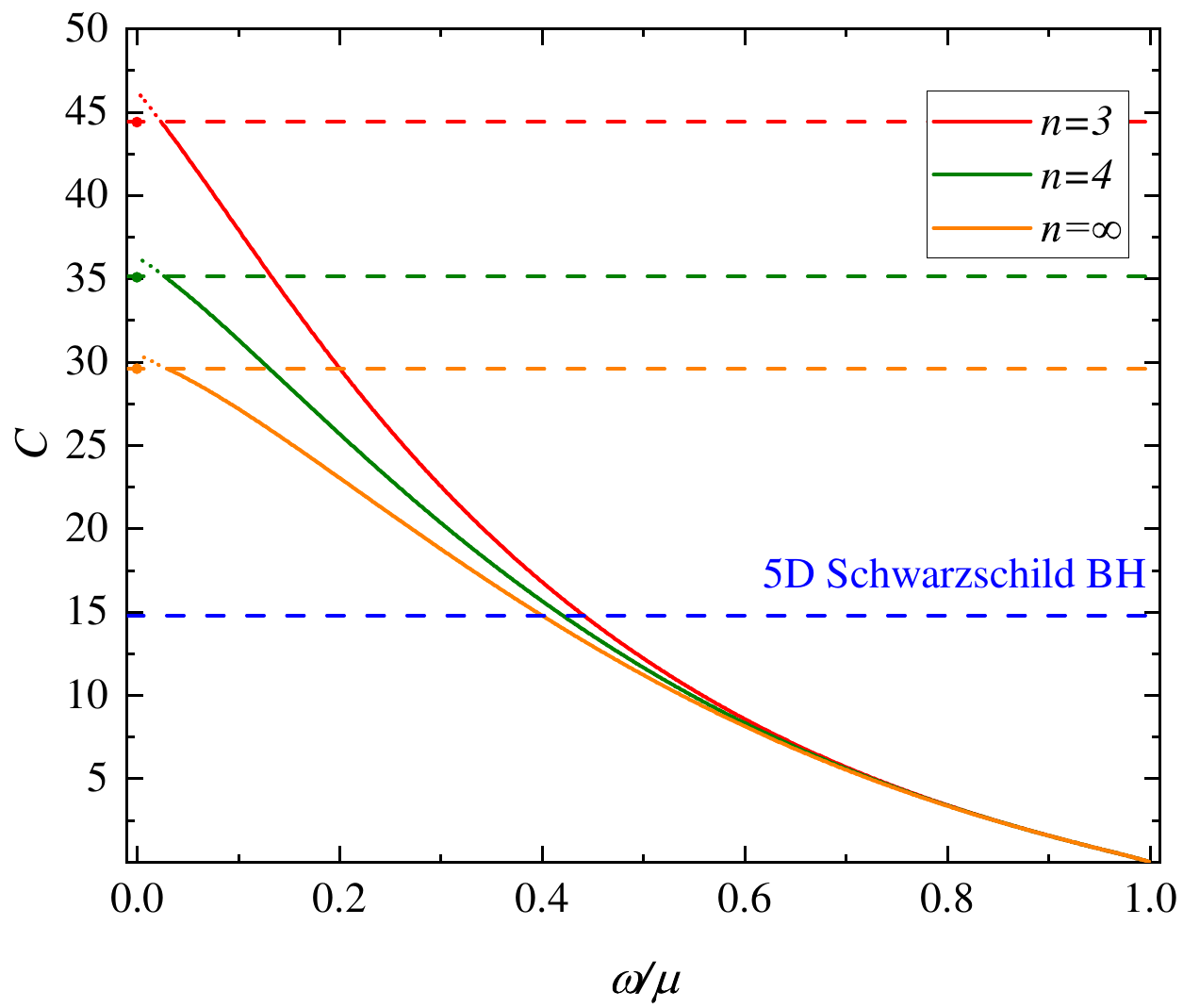}
   \caption{The compactness as a function of frequency for different correction orders $n$ at $\alpha=5$, where the dots mark the compactness of the frozen star solutions at $\omega=0.0002$. The blue dashed line denotes the compactness of the five-dimensional Schwarzschild black hole. The red, green, and orange dashed lines correspond to the extremal black hole compactness for correction orders $n = 2$, $n = 3$, and $n \to \infty$, respectively.}
   	\label{Compactness}
   \end{figure}

We now analyze the energy density distribution and the compactness of the configurations.
Fig.~\ref{Edensity} shows the distribution of energy density as a function of the radial coordinate.
The left panel corresponds to the case $n=2$ at different frequencies, while the right panel displays the energy density for $n=3,4,\infty$ at $\alpha=5$ and $\omega=0.0002$. 
The numerical results indicate that, in the $n=2$ case, the energy density diverges near the center as $\omega \to 0$. 
This behavior is consistent with the divergence of the matter field and the sharp variation of the metric component $1/g_{rr}$ near the central region. 
 In contrast, for the frozen star solutions with $n=3,4,\infty$, the energy density remains finite throughout the entire radial domain and approaches a finite value at the center. 
Moreover, as the correction order increases, the peak value of the central energy density is gradually reduced, indicating that higher-curvature corrections efficiently suppress the divergence of the matter field in the low-frequency regime.

These differences in the energy density distribution further influence the compactness of the configurations. 
In Fig.~\ref{Compactness}, we present the compactness as a function of the frequency for different correction orders at $\alpha=5$. The blue dashed line represents the compactness of the five-dimensional Schwarzschild black hole. The other colored dashed curves indicate the corresponding extremal black hole compactness for each $n$ in the same higher-curvature gravity theory.
Since the Proca field is distributed over the whole space, Proca stars do not possess a rigid surface. 
We therefore define the  effective radius $R_{99}$ as the radius containing $99\%$ of the total mass $\tilde{M}$, i.e.,
\begin{equation}
m(R_{99})=0.99\, \tilde{M} .
\end{equation}
Here, the mass $\tilde{M}$ is given by integrating the matter energy density,
$\tilde{M} = \int dr\, r^2 \rho(r)$.
In four-dimensional spacetime, the compactness of a compact object is conventionally defined as
\begin{equation}
C_{4D}=\frac{\tilde{M}}{R}.
\end{equation}
In higher dimensions, the mass has dimension $[\tilde{M}]=L^{D-3}$. 
Hence, for five-dimensional spacetime, we define the compactness as
\begin{equation}
C_{5D}=\frac{\tilde{M}}{R^{2}}.
\end{equation}
It is noteworthy that, when $\omega \to 0$, the configuration develops a critical horizon at a finite radius $r_c$, and the energy density is primarily concentrated inside this critical horizon. If we still employ $R_{99}$ (defined via $m(R_{99}) = 0.99\tilde{M}$) to compute the compactness of the frozen stars, the resulting curve may exceed the extremal black hole compactness for the corresponding correction order $n$ when $\omega \to 0$. This excess does not imply that the frozen stars are more compact than the extremal black holes. It indicates that $R_{99}$ ceases to be an appropriate measure of the effective radius once the critical surface at $r_c$ forms. The dots in the figure show the compactness of the frozen-star configurations at $\omega = 0.0002$, evaluated using the critical radius $r_c$. These dots coincide with the extremal black holes for each correction order 
$n$, indicating that the compactness of frozen stars is consistent with their corresponding extremal black hole limits.
For solutions with $n\geq 3$, the configurations become increasingly compact as the frequency decreases. And the $\omega\to 0$ corresponds to the most compact configurations for a given correction order. In addition, for frozen-star solutions, the compactness decreases as the correction order $n$ increases.

      \subsection{ Light Rings} 
   \begin{figure}[!htbp]
   	\centering
   	\includegraphics[width=0.45\textwidth]{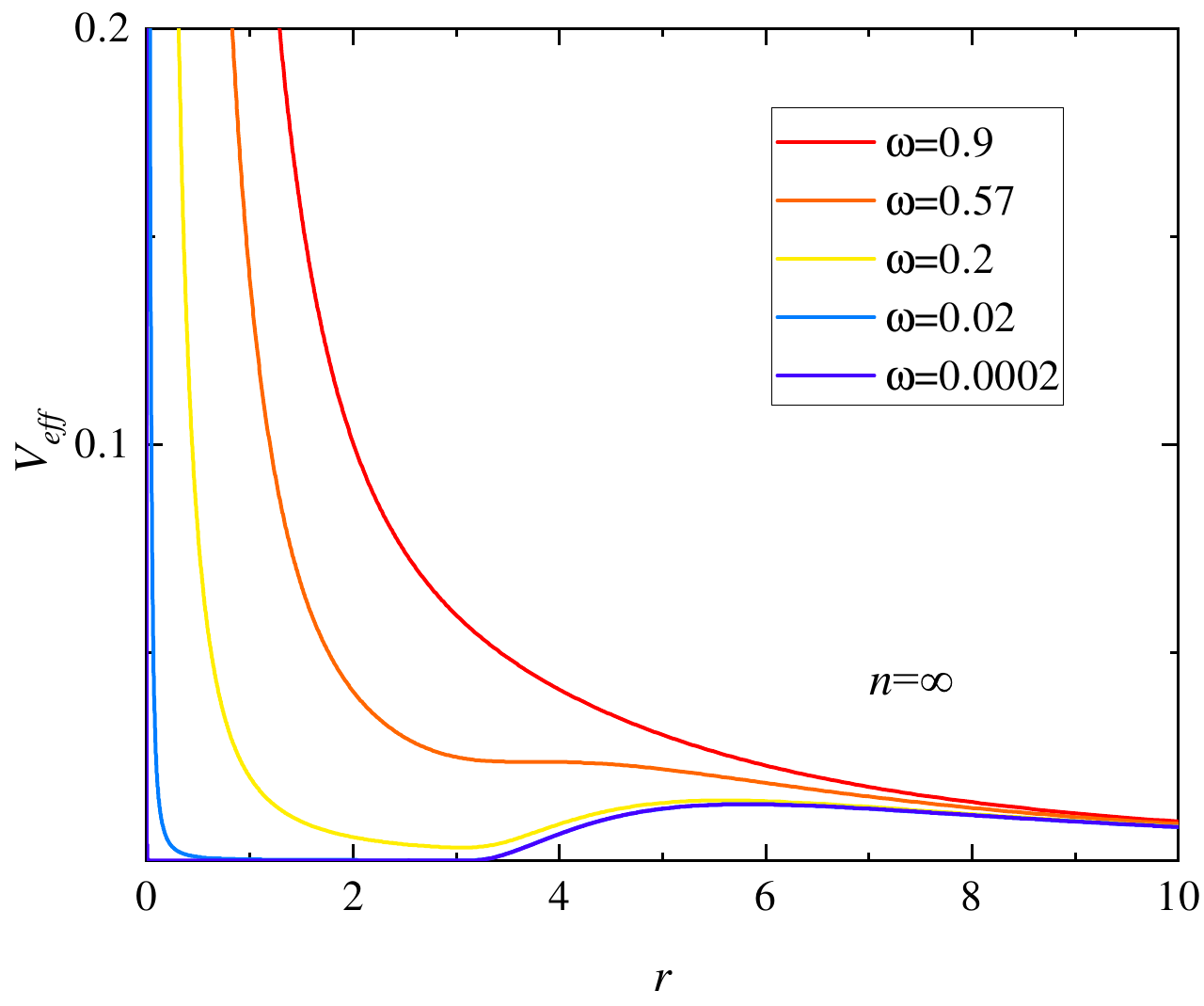}
    \includegraphics[width=0.45\textwidth]{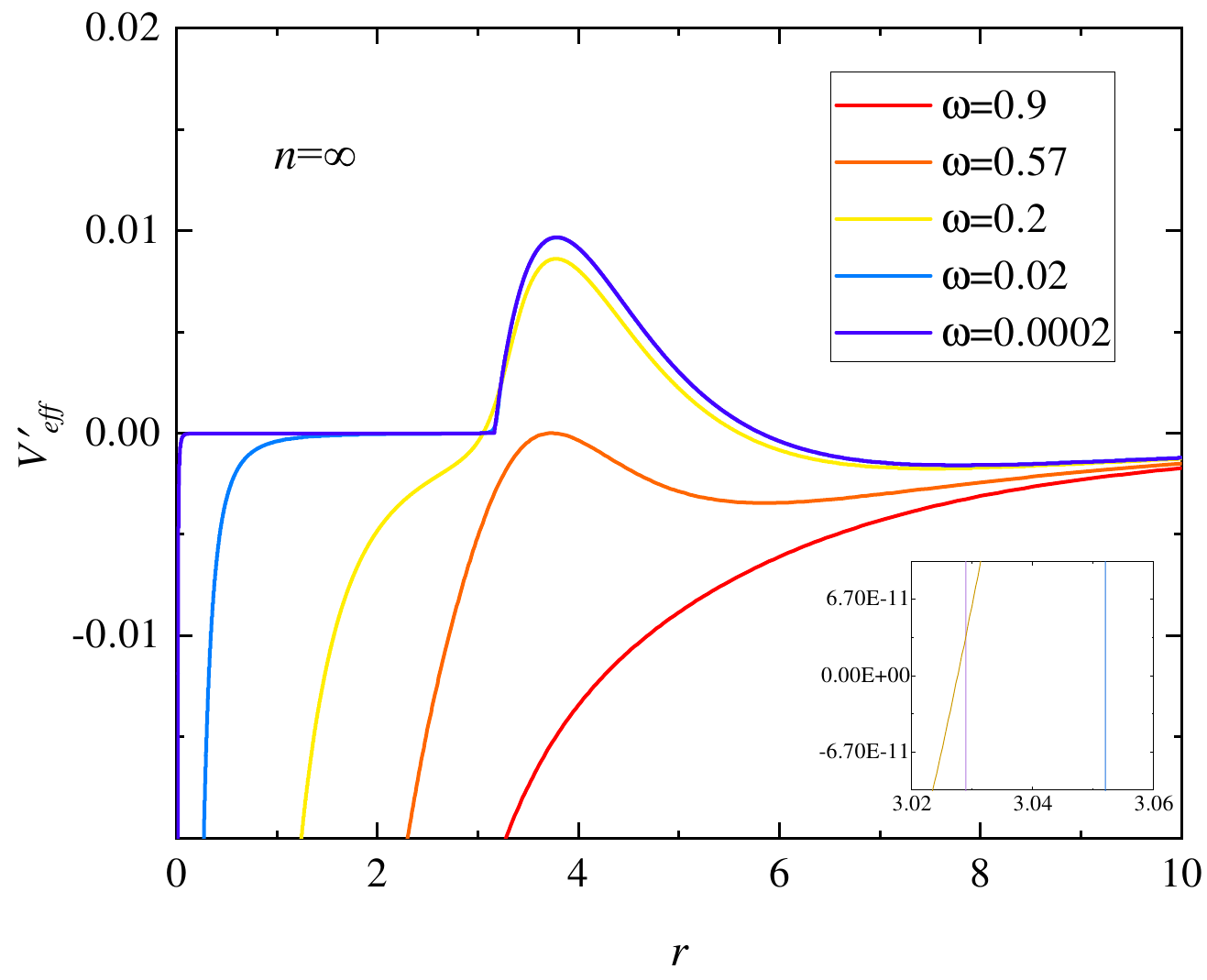}
   	\caption{The effective potential $V_{eff}$ (left panel) and its first derivative $V'_{eff}$ (right panel) as functions of the radial coordinate for $n=\infty$.}
   	\label{effective}
   \end{figure} 
In general relativity, gravity is described by the curvature of spacetime, and massless particles travel along its null geodesics.
In strong gravitational fields, photon trajectories can exhibit circular orbits, known as light rings~\cite{Cardoso:2014sna}. In a static and spherically symmetric spacetime, photon motion is governed by the null geodesic equation~\cite{Thorne:1973}
\begin{equation}\label{equ28}
g_{\mu\nu}\dot{x}^\mu\dot{x}^\nu = 0 ,
\end{equation}
where the dot denotes differentiation with respect to the affine parameter $\lambda$ along the geodesic. 
Due to the existence of Killing vectors associated with time translations and spatial rotations, two conserved quantities arise, namely the photon energy $E$ and angular momentum $L$~\cite{chandrasekhar1998mathematical}, 
\begin{equation}\label{equ29}
E = - g_{tt}\dot{t}, \qquad L = r^2\dot{\phi}.
\end{equation}
Without loss of generality, the motion can be restricted to an equatorial plane.

Bring these conserved quantities Eq.~(\ref{equ29}) in the null geodesic equation  Eq.~(\ref{equ28}) yields the radial equation of motion~\cite{Kar:2023dko,
Rosa:2022tfv,
Herdeiro:2021lwl},
\begin{equation}
\dot{r}^2 +\frac{L^2}{g_{tt} g_{rr}}\left(\frac{1}{b^2}+\frac{g_{tt}}{r^2}\right)=0,
\end{equation}
where $b \equiv L/E$ is the impact parameter. 
With the metric ansatz of Eq.~(\ref{equ3}), the equation becomes
\begin{equation}
\dot{r}^2 +\frac{L^2}{\sigma(r)^2}\left(-\frac{1}{b^2}+\frac{\sigma(r)^2N(r)}{r^2}\right)=0.
\end{equation}
It is therefore convenient to define the effective potential
\begin{equation}
V_{\rm eff}(r)=\frac{\sigma(r)^2N(r)}{r^2}.
\end{equation}
Light rings correspond to circular photon orbits. 
Their existence requires that both the radial velocity and its first derivative vanish. 
As a result, the light ring radius $R_{\rm lr}$ is determined by the extremum of the effective potential~\cite{Cardoso:2021sip},
\begin{equation}
\frac{dV_{\rm eff}}{dr}\bigg|_{R_{\rm lr}} = 0 .
\label{lightring}
\end{equation}

The stability of a light ring is governed by the second derivative of the effective potential~\cite{
Cunha:2022gde,
Cunha:2017wao}. 
If the effective potential has a local maximum, i.e.\ $V^{''}_{\rm eff}(R_{lr})<0$, the corresponding photon orbit is unstable, and small perturbations cause photons to depart from the circular trajectory. 
Conversely, if $V^{''}_{\rm eff}(R_{lr})>0$, the effective potential has a local minimum and the circular orbit is stable, with perturbed photons oscillating around the equilibrium radius.

    \begin{figure}[!htbp]
   	\centering
   	\includegraphics[width=0.45\textwidth]{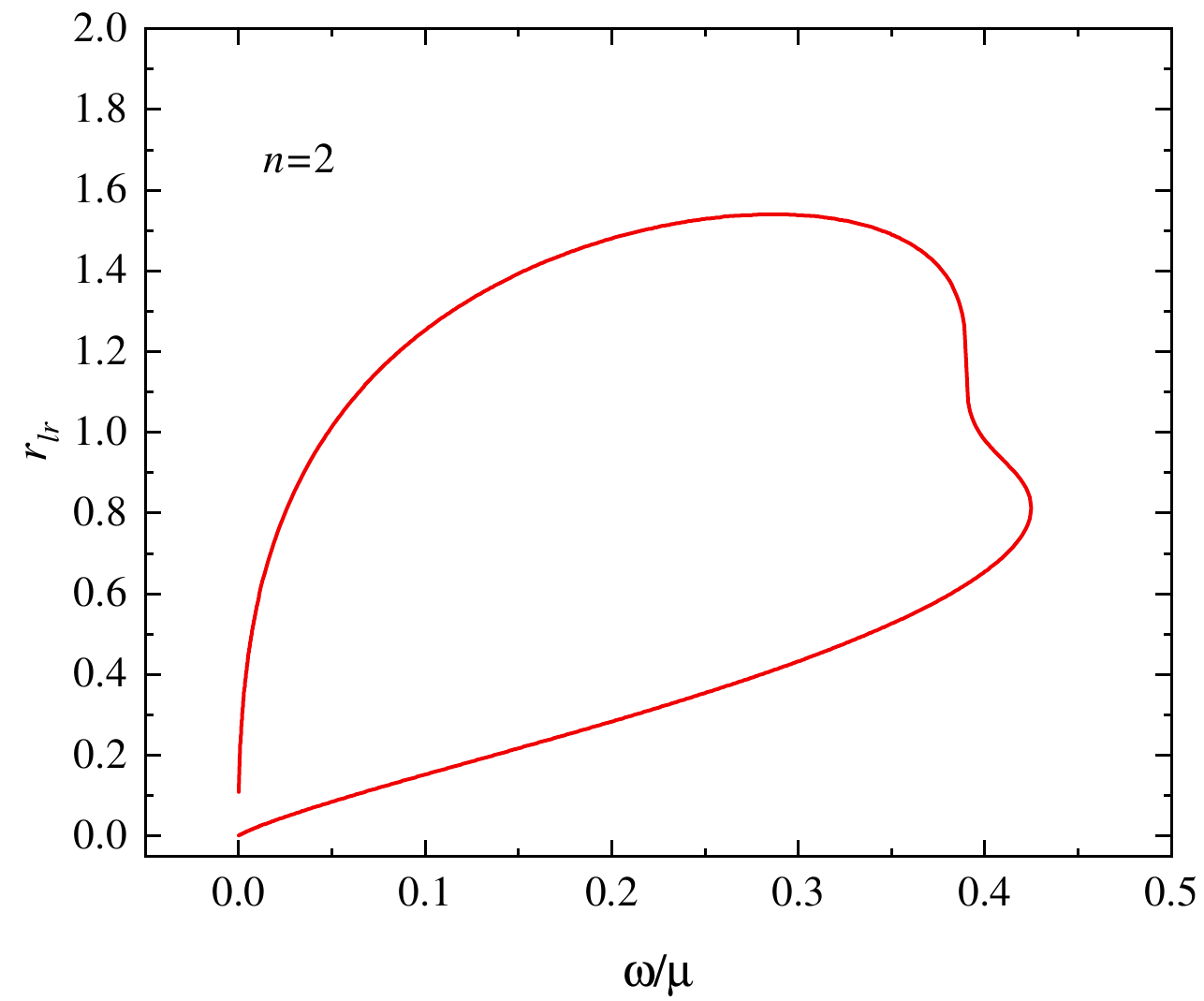}
    \includegraphics[width=0.45\textwidth]{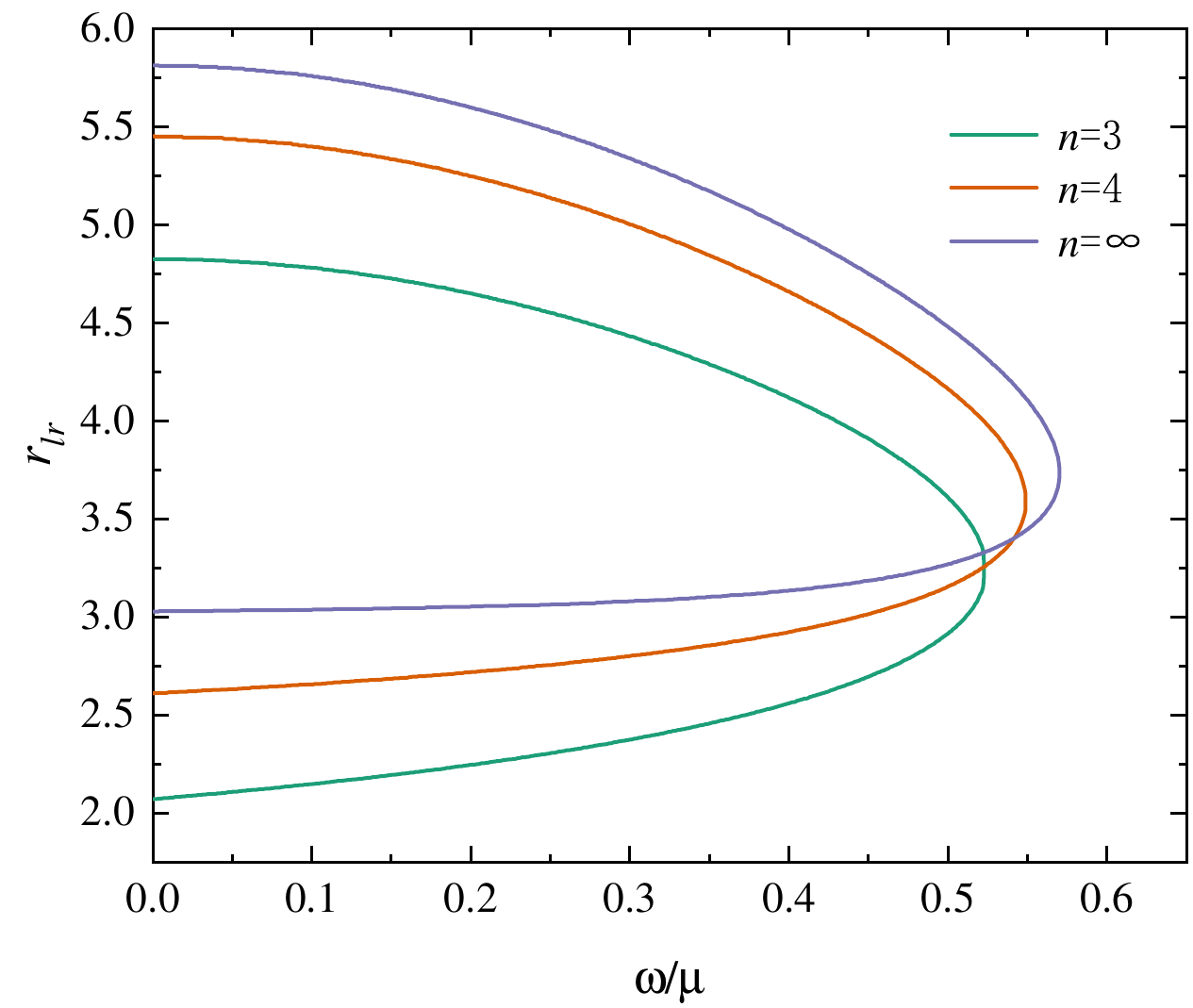}
   	\caption{ The light ring positions as a function of the frequency for $n=2$ (left panel) and $n=3,4,\infty$ (right panel).}
   	\label{light}
   \end{figure} 
   
In Fig.~\ref{effective}, we present the effective potential $V_{\rm eff}$ and its first derivative $V'_{\rm eff}$ as functions of the radial coordinate for the infinite-order curvature correction, at different values of the frequency. 
The existence of light rings is closely related to the compactness of the configuration. Only when the matter distribution is sufficiently concentrated within a finite radial region can the effective potential develop extrema that support circular photon orbits.
At high frequencies, where the compactness is relatively low, $V_{\rm eff}$ varies monotonically across the entire radial domain without any extremum and no light ring exists. 
As the frequency decreases, the configuration becomes more compact and the effective potential develops a single extremum, corresponding to the emergence of a light ring. 
Upon further lowering the frequency, two extrema appear in $V_{\rm eff}$, giving rise to a double light ring structure. 
For $\omega = 0.0002$, the effective potential clearly exhibits two extrema, indicating that the frozen star solution possesses two light rings. 
The inner light ring is stable, while the outer one is unstable.
To further investigate the effect of the curvature correction order on the light-ring structure, we show in Fig.~\ref{light} the locations of the light rings as functions of the frequency for $n=2,3,4,\infty$. 
The results indicate that, as the correction order $n$ increases, both the inner and outer light rings of frozen star configurations shift toward larger radial distances.


     \section{Conclusion} \label{sec5}
In this work, we have investigated five-dimensional static and spherically symmetric Proca star solutions in gravity theories with higher-curvature corrections. 
Our numerical results show that, for $n=1$, the model reduces to Proca stars in five-dimensional Einstein gravity~\cite{Hartmann:2010pm,
Hartmann:2012gw,
Blazquez-Salcedo:2019qrz}. 
In this case, the allowed frequency range of the solutions is finite and no frozen star solutions are found. 
For the $n=2$ (Gauss--Bonnet case), the domain of existence is extended and configurations with frequencies approaching zero appear. 
However, in the $\omega \to 0$ limit, both the matter fields and the energy density diverge near the center. 
By contrast, for higher-order curvature corrections with $n \geq 3$, the matter fields and the energy density remain finite throughout the entire spacetime.

For $n \geq 3$, we find a class of frozen star solutions that do not possess an event horizon. Interestingly, frozen star solutions can be obtained already at finite order in the curvature corrections, without the necessity for infinite-order terms.
As the correction order $n$ decreases, the frozen star configurations become more compact. Notably, the compactness of frozen stars is identical to that of their corresponding extremal black holes. The matter fields and energy density are mainly concentrated inside the critical horizon. 
Outside this critical horizon, the metric of the frozen star solutions coincides with that of extremal black hole solutions with the same mass and coupling constant. 
Consequently, for a distant observer at infinity, frozen stars are indistinguishable from extremal black holes based solely on exterior observations. 
Unlike black holes, however, frozen stars do not contain curvature singularities that invalidate physical laws, nor do they possess an event horizon. 
Therefore, as a potential alternative to black holes, frozen stars not only provide a possible resolution to the black hole information loss problem associated with the presence of an event horizon, but also offer new insights toward the construction of a self-consistent theory of quantum gravity.

Future work will focus on the perturbative stability of frozen star solutions by analyzing their dynamical evolution under perturbations to assess their astrophysical viability. Furthermore, In classical GR, the tidal Love numbers of black holes vanish~\cite{Fang:2005qq}, which is the key feature distinguishing them from other compact objects~\cite{Cardoso:2017cfl,
Sennett:2017etc}. Therefore, it is also important to investigate the tidal response of frozen stars.

	\section*{ACKNOWLEDGEMENTS}

This work is supported by the National Natural Science Foundation of China (Grant
No. 12275110 and No. 12247101) and the National Key Research and Development Program of
China (Grant No. 2022YFC2204101 and 2020YFC2201503).

\end{document}